\documentclass[12pt]{article}
\usepackage{amsmath}
\usepackage{amsfonts}
\usepackage{amssymb}
\usepackage{indentfirst}
\def\llsymbol#1{\@llsymbol{\@nameuse{c@#1}}}
\def\@llsymbol#1{\ifcase#1\or {}\or {'}\or {''}\or {'''}\or
   {''''}\or {'''''}\or  \else\@ctrerr\fi\relaz}
\newcounter{contador}
\newcommand{\letra}{
   \stepcounter{equation}
   \setcounter{contador}{\value{equation}}
   \setcounter{equation}{0}
   \renewcommand{\theequation}{\thecontador\alph{equation}}}
\newcommand{\antiletra}{
   \renewcommand{\theequation}{\arabic{equation}}
   \setcounter{equation}{\value{contador}}}

\begin{document}
\begin{center}{\Large \bf Generalized spheroidal wave equation and limiting cases }
\vskip 0.4cm {B. D. Bonorino Figueiredo} \\
 Instituto de Cosmologia, Relatividade e Astrof\'isica (ICRA-BR)\\
{Centro Brasileiro de Pesquisas F\'{\i}sicas (CBPF)\\
 Rua Dr. Xavier Sigaud, 150 - 22290-180 - Rio de Janeiro, RJ, Brasil}
\end{center}
%

%
\begin{abstract}
\noindent
We find sets of solutions to the generalized spheroidal wave
equation (GSWE) or, equivalently, to the confluent Heun equation.
Each set is constituted by three solutions, one given by a series
of ascending powers of the independent variable,
and the others by series of regular and  irregular
confluent hypergeometric
functions. For a fixed set, the solutions converge over
different regions
of the complex plane but present series coefficients
proportional to
each other. These solutions for the GSWE afford solutions
to a double-confluent Heun equation by a taking-limit process due
to Leaver. Another procedure, called Whittaker-Ince limit,
provides solutions in series of powers
and Bessel functions for two other equations with
a different type of singularity at infinity. In addition,
new solutions are obtained for the Whittaker-Hill
and Mathieu equations
by considering these as special cases
of both the confluent and double-confluent Heun equations.
In particular, we find that each of the Lindmann-Stieltjes solutions
for the Mathieu equation is associated with two expansions in
series of Bessel functions. We also discuss a set of solutions
in series of hypergeometric and confluent hypergeometric
functions for the GSWE and use their Leaver limits to obtain
infinite-series solutions for the Schr\"odinger equation
with an  asymmetric double-Morse potential.
Finally, the possibility of extending the solutions
of the GSWE to the general Heun equation is briefly discussed.


%
\end{abstract}
%
%
\section*{I. Introduction}
\indent
We study sets of solutions for a generalized spheroidal
wave equation (GSWE) aiming at obtaining solutions
for other differential equations which are related to the GSWE by
taking-limit processes. Although the three solutions which
constitute a fixed set converge over different  regions of the complex
plane, they have series coefficients proportional to one another
what means that, under certain circumstances, the three solutions
imply a unique characteristic equation resulting from the
three-term recurrence relations for the coefficients.
Solutions for a double-confluent Heun
equation (DCHE) are generated from the ones for the GSWE
by using of a limit process devised by Leaver \cite{leaver1}
and, therefore, without solving directly the DCHE.
On the other hand, by applying to the GSWE and DCHE another limit
due to Whittaker and Ince \cite{humbert,ince,eu},
we obtain two differential
equations with a different type of singularity at infinity
whose solutions are obtained from the ones for GSWE
and DCHE by taking the Whittaker-Ince limit, again
without integrating
directly the equations. Thus, by means of the
Leaver and Whittaker-Ince limits, the GSWE afford solutions for
three other different equations.

Furthermore, new
solutions for the Whittaker-Hill and Mathieu equations are
obtained from the fact that these equations are particular cases
of both the GSWE and the DCHE \cite{decarreau1}.
From one side, this gives two types of solutions
for the Whittaker-Hill
equation (WHE) and, from the other side, three
types for the Mathieu
equation in virtue that this one is also
an instance of the Whittaker-Ince limit of the GSWE.

In summary, solutions for the GSWE generate solutions for
five other equations which in most of the cases have
been studied separately and without any connection with the
GSWE, namely: DCHE, Ince's limits of
the GSWE and DCHE, and the Mathieu and Whittaker-Hill
equations. Some solutions we shall find are already known,
but here they are associated with
two other solutions possessing different mathematical
properties, since we are dealing with sets containing three
solutions, as mentioned in the first paragraph
(the known solutions result from a known solution
for the GSWE given by a series of ascending powers of
the independent variable).

In addition to the solutions for
the GSWE and its limiting and particular cases, we
consider two physical problems.
One is given by the Schr\"odinger equation for
asymmetric double-Morse potentials concerning
quantum spin systems \cite{zaslavskii,ulyanov}. For
quasi-exactly solvable potentials \cite{usheveridze1,usheveridze2},
an earlier attempt to solve this equation
had been unsuccessful \cite{eu2}. Here we show that the
solution of this problem requires the use of two
different types of solutions for the DCHE, both of them
discussed in this article. The other problem is given by
the Teukolsky equations for gravitational backgrounds of black holes,
which afford a connection among three equations:
general Heun equation, GSWE and DCDE \cite{mano4,mano5}.
However, this problem is regarded only at the extent it
helps to find solutions for the Heun equation and its
confluent cases.

After these general considerations, let us examine each of
these six equations and how they are connected,
discuss some features of the solutions and outline
the structure of the paper.

%
{\it The generalized spheroidal wave equation (GSWE)}.
For the GSWE we adopt the form used by Leaver \cite{leaver1},
\begin{eqnarray}
\label{gswe}
z(z-z_{0})\frac{d^{2}U}{dz^{2}}+(B_{1}+B_{2}z)
\frac{dU}{dz}+
\left[B_{3}-2\eta
\omega(z-z_{0})+\omega^{2}z(z-z_{0})\right]U=0, \ (\omega\neq0)
\end{eqnarray}
where $B_{i}$, $\eta$ and $\omega$ are constants
and $z=0$ and $z=z_{0}$ are regular singular
points with indicial exponents ($0,1+B_{1}/z_{0}$)
and ($0,1-B_{2}-B_{1}/z_{0}$), respectively, that is,
\begin{eqnarray}\label{frobenius}
\begin{array}{ll}
\displaystyle \lim_{z\rightarrow  0}U(z)\sim1  &
\ \mbox{or}\ \  \displaystyle \lim_{z\rightarrow0}U(z)\sim z^{1+(B_{1}/z_{0})},
\vspace{3mm}\\
\displaystyle\lim_{z\rightarrow  z_{0}}U(z)\sim 1 &
\ \mbox{or}\ \  \displaystyle\lim_{z\rightarrow  z_{0}}U(z)
\sim (z-z_{0})^{1-B_{2}-(B_{1}/z_{0})}.
\end{array}
\end{eqnarray}
At $z=\infty$, which is an irregular singularity, the
behavior inferred from the
normal Thom\'e solutions \cite{leaver1,olver} is
\begin{eqnarray}\label{thome1}
\lim_{z\rightarrow  \infty}U(z)\sim e^{\pm i\omega z}z^{\mp i\eta-(B_{2}/2)}.
\end{eqnarray}
Eq. (\ref{gswe}) is in a form appropriate for the
study of the Teukolsky
equations for the Kerr metric since, in the upper limit
for the rotation parameter, $z_{0}$ vanishes and the
equation reduces to an DCHE. In fact,
Eq. (\ref{gswe}) is equivalent to the
confluent Heun equation \cite{decarreau1,decarreau2,ronveaux} and,
when $\eta=0$, it reduces to the so-called ordinary spheroidal
wave equation  \cite{wilson}.

%

%

{\it The double-confluent Heun equation (DCHE)}.
By setting $z_{0}=0$ in Eq. (\ref{gswe}), Leaver
obtained a DCHE with five parameters \cite{leaver1},
namely,
\begin{eqnarray}\label{dche}
z^{2}\frac{d^{2}U}{dz^{2}}+
\left(B_{1}+B_{2}z\right)\frac{dU}{dz}+
\left(B_{3}-2\eta \omega z+\omega^{2}z^{2}\right)U=0,
\ \left(B_{1}\neq 0, \  \omega\neq 0\right),
\end{eqnarray}
where $z=0$ and $z=\infty$ are both irregular
singularities. At $z=\infty$ the behavior of the solutions is again
given by Eq. (\ref{thome1}), while at $z=0$ the normal
Thom\'e solutions affords
\begin{eqnarray}\label{thome}
\displaystyle\lim_{z\rightarrow  0}U(z)\sim 1,\ \mbox{or}\
 \displaystyle\lim_{z\rightarrow  0}U(z)\sim
 e^{B_{1}/ z}z^{2-B_{2}}.
\end{eqnarray}
The values $B_{1}=0$ and/or
$\omega=0$ (for $\eta$ fixed)
are excluded because: if $B_{1}=\omega=0$ we
have the Euler equation; if $B_{1}=0$ and
$\omega\neq 0$ or if $B_{1}\neq 0$
and $\omega=0$, the equation degenerates
into a confluent hypergeometric equation
(Appendix A of \cite{eu}).

Here the Leaver limit $z_{0}\rightarrow 0$ is
used to obtain solutions for the DCHE
(\ref{dche}) from solutions of the GSWE. Furthermore, the Leaver
form (\ref{dche}) for the DCHE has also the  advantage
of admitting the Whittaker-Ince limit, in opposition to other forms which
have only four parameters, as in Refs.
\cite{decarreau1,decarreau2,schmidt}.


%
{\it The Whittaker-Ince limit of the GSWE}.
We define the Whittaker-Ince limit of
the GSWE and DCHE as
\begin{eqnarray}\label{ince}
\omega\rightarrow 0, \ \
\eta\rightarrow
\infty, \ \mbox{such that }\  \ 2\eta \omega =-q,\ \
(\mbox{Whittaker-Ince limit})
\end{eqnarray}
where $q$ is a constant. In a previous article \cite{eu}
we have referred to the above as 'Ince limit', but
according to Humbert \cite{humbert},
prior to Ince \cite{ince}, Whittaker had used
such procedure to get the Mathieu equation from
the Whittaker-Hill one. In any case, the limit (\ref{ince})
is a generalization of the limit used by Whittaker and Ince.

The Whittaker-Ince limit of the GSWE is
\begin{eqnarray}
\label{incegswe}
z(z-z_{0})\frac{d^{2}U}{dz^{2}}+(B_{1}+B_{2}z)
\frac{dU}{dz}+
\left[B_{3}+q(z-z_{0})\right]U=0,\ (q\neq0)
\end{eqnarray}
(if $q=0$ this equation can be transformed into a hypergeometric
equation). At the regular singular points $z=0$
and $z=z_{0}$ the behaviors of the
solutions are formally the same as those for the GSWE, but
at the irregular singular point $z=\infty$, from the subnormal
Thom\'e solutions \cite{olver} we get
\begin{eqnarray}\label{thome2}
\lim_{z\rightarrow  \infty}U(z)\sim
e^{\pm 2i\sqrt{qz}}z^{(1/4)-(B_{2}/2)}.
\end{eqnarray}
in contrast with the behavior (\ref{thome1}) of original GSWE
(normal Thom\'e solutions).

Mignemi \cite{mignemi} and
Malmendier \cite{malmendier} have found
a particular case ($B_{2}=2B_{1}$, $\ z_{0}=-1$) of Eq.
(\ref{incegswe}) for a wave equation resulting from
the Laplace-Beltrami operator for a scalar field
on the Eguchi-Hanson
space \cite{eguchi}, but they
did not succeed in deriving solutions from solutions
of the GSWE. Here and in Ref. \cite{eu} this
goal is attained by starting with solutions for the
GSWE which admit both the Leaver and the Whittaker-Ince
limits. One can check that this is not possible for some
other solutions as the Hylleraas and Jaff\'e ones \cite{leaver1},
for example.
%


{\it The Whittaker-Ince limit of the DCHE}. The
Whittaker-Ince limit of the
DCHE is the equation
\begin{eqnarray}
\label{incedche}
z^2\frac{d^{2}U}{dz^{2}}+(B_{1}+B_{2}z)
\frac{dU}{dz}+
\left(B_{3}+qz\right)U=0,\ (q\neq0, \ B_{1}\neq 0)
\end{eqnarray}
where $z=0$ and $z=\infty$ are irregular singularities
as in the DCHE (\ref{dche}). At the irregular singular point
$z=0$ the solutions behave again according to Eq. (\ref{thome}),
but at infinity they admit the behavior given by the subnormal
Thom\'e solutions, Eq. (\ref{thome2}).

Eq. (\ref{incedche}) is not obtainable from a DCHE
with four parameters. In fact,
the scale transformation in the variable $z$ used to
pass from five into four parameters \cite{decarreau1,decarreau2}
becomes meaningless
when $\omega=0$ in Eq. (\ref{dche}). So, the Leaver
form (\ref{dche}) is valuable since it leads
to solutions for Eq. (\ref{incedche}) which rules, for example, the
radial dependence of the Schr\"odinger equation
for the scattering of low-energy
ions by polarizable neutral atoms with an induced
quadrupole momentum \cite{eu}.

In Eq. (\ref{incedche}) $q=0$ and/or $B_{1}=0$
were excluded
because in these cases the equation can be
transformed into a confluent hypergeometric
equation if $q=0$ and $B_{1}\neq 0$, into a Bessel equation
if $q\neq0$ and $B_{1}= 0$, and into an Euler equation
if $q=B_{1}= 0$ \cite{eu} .

The previous equations -- (\ref{gswe}), (\ref{dche}), (\ref{incegswe})
and (\ref{incedche}) -- can also be distinguished
by the rank or {\it species} of the singularity at $z=\infty$
\cite{decarreau1,decarreau2}. For the
GSWE (\ref{gswe}) and the DCHE (\ref{dche})
the rank is $1$ ({\it species $2$}), while for the
Whittaker-Ince
limits (\ref{incegswe}) and (\ref{incedche}) the rank is $1/2$
({\it species} 1). However, here this classification is not relevant.
To get the solutions, more important are the
concepts of Leaver and Whittaker-Ince limits, and the information
on whether the solutions exhibit the behavior
predicted by either the normal or the subnormal
Thom\'e solutions.

Notice that Eq. (\ref{incedche}) is also the Leaver limit of
Eq. (\ref{incegswe}) and, consequently, we have two
ways to generate its solutions, as indicated in the following
scheme:
\begin{eqnarray}\label{scheme}
\begin{array}{clc}
\boxed{\mbox{GSWE (1)}}&\longrightarrow\mbox{\ \ Leaver limit}&
\longrightarrow\boxed{\mbox{DCHE (4)}}\\
 \hspace{0.9cm} \downarrow&                   &  \hspace{1.7cm} \downarrow\\
\mbox{Whittaker-Ince limit}&                        & \ \ \ \ \ \
\mbox{Whittaker-Ince limit}\\
\hspace{0.9cm} \downarrow&                    & \hspace{1.7cm}  \downarrow\\
\boxed{\ \mbox{Eq. (7) }}&\longrightarrow\mbox{\ \ Leaver limit}&
\longrightarrow\boxed{\ \mbox{Eq. (9) }}.
\end{array}
\end{eqnarray}
A preliminary study of solutions for the Heun equation indicates
that above diagram can be improved by adding the connection
Heun equation $\to$ GSWE (see Appendix B).

%

{\it Whittaker-Hill and Mathieu equations}.
The Whittaker-Hill and Mathieu equations are
particular cases of both the
GSWE (\ref{gswe}) and the DCHE (\ref{dche}),
but the Mathieu equation is also
a special instance of the Whittaker-Ince limit of the GSWE.
Thence, there are two types of solutions for the Whittaker-Hill
equation (WHE) and three types for the Mathieu
equation, as stated above. This result is due to
Decarreau, Maroni and Robert \cite{decarreau1}
but it is necessary to study the properties of
the solutions obtained by each of these approaches, by
writing such solutions explicitly
and comparing them.

To avoid confusion in the notation, we write the Whittaker-Hill
and Mathieu equations in a form independent of
the parameters $B_{i}$, $\eta$ or $\omega$ which
appear in the other equations. We choose the
trigonometric (hyperbolic) form  because this is
useful to investigate the parity and periodicity properties
of such solutions. For the WHE we employ
the form \cite{ince,arscott}
\begin{eqnarray}\label{whe}
\frac{d^2W}{du^2}+\kappa^2\left[\vartheta-\frac{1}{8}\xi^{2}
-(p+1)\xi\cos(2\kappa u)+
\frac{1}{8}\xi^{2}\cos(4\kappa u)\right]W=0, \ \ (\mbox{WHE})
\end{eqnarray}
where we have written $\vartheta$, instead of the
usual $\eta$, because $\eta$ has already appeared in
the GSWE and DCHE. If $u$ is a real variable, this
equation represents the WHE when $\kappa=1$ and
the modified WHE when $\kappa=i$. The WHE
(\ref{whe}) is converted into a GSWE or DCHE in
Secs. II B or III B, respectively.
For the Mathieu equation we use the
form \cite{McLachlan}
\begin{eqnarray}\label{mathieu}
\frac{d^2W}{du^2}+\sigma^2[a-2k^2\cos(2\sigma u)]W=0,
 \ \ (\mbox{Mathieu equation})
\end{eqnarray}
where  $\sigma=1$ or $\sigma=i$ for the Mathieu or
modified Mathieu equation, respectively. Note
that there are two ways to get Eq. (\ref{mathieu})
from the WHE (\ref{whe}). First, for $p=-1$ and
$\kappa=\sigma/2$, the WHE (\ref{whe}) reduces
to the Mathieu equation (\ref{mathieu}). Second,
the Mathieu equation (\ref{mathieu}) also follows
from the WHE (\ref{whe}) by means of the original
Whittaker-Ince limit \cite{humbert,ince,ince2}: $\xi\rightarrow 0$,
$p\rightarrow \infty$ so that $p\xi=2k^2$,
$\kappa=\sigma$ and $\vartheta=a$. This
second way is equivalent
to consider the Mathieu equation as a particular case of
the Whittaker-Ince limit (\ref{incegswe}) of the GSWE.
The Mathieu equation (\ref{mathieu}) is transformed into
a GSWE, a DCHE or the Ince limit of the GSWE in Secs. II C,
III C or IV B, respectively.
%


In this paper, starting from sets formed by three solutions for the GSWE,
we follow the scheme (\ref{scheme}) to generate solutions
for the other equations. In each set, one solution
is given by a Barber-Hass\'e expansion in power
series \cite{leaver1,barber} while
the others are given by expansions in series of
regular and irregular hypergeometric
functions. In a fixed set, the solutions converge over
different regions but are collected together because
their series coefficients are proportional to each other.
The construction of sets of solutions with these features
is in part suggested by study of the Mathieu
equations \cite{McLachlan}, but it follows
as well  from the study of wave equations in general relativity
where sometimes is necessary to match solutions
convergent over different domains
\cite{leaver1,eu2,otchik1,mano1,mano2,mano3}.

In the context of the GSWE, a initial
set, as the one given in Eq. (\ref{frobenius1a}), leads to the
others by means of transformation rules
coming from variable substitutions which
preserve the form of the GSWE but modify their parameters
and/or arguments. The transformation rules for the GSWE,
written in Eqs. (\ref{rule}), allow us
to generate solutions with all the expected behaviors
at the singular points. These features are transferred
to the solutions of the limiting equations and
their particular cases. This is illustrated by the results
of Sections II B and IV B, where we find that
each set of solutions for the Whittaker-Hill and
Mathieu equations, respectively, possesses
different properties of parity and /or periodicity.

On the other hand, the convergence of the series
depends on whether the parameters of the differential equation
do or do not satisfy a relationship, known as characteristic equation,
which results from the three-term recurrence relations
for the series coefficients. In fact, there are two
cases to be considered. First, if
there is some arbitrary constant in the differential equation,
the characteristic equation can be satisfied by using it
to determine that constant in terms of the others and in such
case the series converges. If there is no disposable constant,
the characteristic equation in general cannot be satisfied
and then the series diverges. In this second case,
the convergence is assured by inserting into certain
solutions a parameter $\nu$, called phase or
characteristic parameter,  which plays the role of the arbitrary
constant and must be adjusted so as to validate the characteristic
equation. Excepting section VI, all the other sections of the
present article deal with solutions without characteristic
parameter and, consequently,
assume the existence of an arbitrary constant in the differential
equations.

In section II A, we study  the Barber-Hass\'e solutions and the
expansions in series of confluent hypergeometric functions
for the GSWE. In section II B  these solutions are particularized
for the Whittaker-Hill equation and in Sec. II C for the
Mathieu equation.
In each set of solutions for the WHE, the solution resulting
from the Barber-Hass\'e one turns out to be a solution
already found by Arscott \cite{arscott}, but the
others seem to be new.

In Sec. III A the Leaver limit $z_{0}\rightarrow 0$ is
used to derive solutions
for the DCHE from solutions for the GSWE. The solutions
ensuing from the  Barber-Hass\'e  solutions include as particular
cases the ones which have been investigated by Decareau,
Maroni and Robert \cite{decarreau2}
and also by Schmidt and Wolf \cite{schmidt}
for a DCHE with four parameters, but to each of these
now correspond two expansions in series of
confluent hypergeometric functions. These solutions are
used to find finite-series solutions for the
Schr\"odinger equation
with an asymmetric double-Morse potential.
There are infinite-series solutions but these are not
regular at the singular points of the equation.
The WHE and the Mathieu equations as particular cases
of the DCHE are discussed in Secs.
III B  and III C, respectively.

In Sec. IV  we deal with the Whittaker-Ince limit of the GSWE and
find that the expansions in series of confluent
hypergeometric functions reduce to expansions in
series of Bessel functions. In the special case of
the Mathieu equation, we recover
some solutions found by Lindmann  and Stieltjes
\cite{lindemann,watson},
each one coupled with two expansions
in series of Bessel functions. Similarly, in Sec.
V we establish the Ince limits for the solutions of the DCHE.

In Sec. VI we consider some solutions with a phase
parameter for the case in which {\bf there is no free} constant
in the GSWE. Each set of solutions for the GSWE contains
two expansions
in series of hypergeometric functions and two
in series of confluent hypergeometric functions.
These solutions also admit of both the Leaver
and Whittaker-Ince limits. The solutions for the DCHE
are used to find regular infinite-series solutions for the Schr\"odinger
equation with the asymmetric double-Morse potential
mentioned above. For this it is necessary to match two
solutions converging over different ranges of the
independent variable but having a common region
of convergence. The results is that, for this quasi-exactly potential,
the entire spectrum of energies may be determined by solving
a characteristic equation given by the sum of two infinite
continued fractions.

Section VII contains some concluding remarks, while Appendix
A presents an alternative derivation of the expansions in series of
Bessel functions for the Whittaker-Ince limit of the GSWE and Appendix B
discusses some connections between solutions of the general
Heun equation and generalized spheroidal
equation.
%
%
%
%
%
%
\section*{II. Solutions to the generalized spheroidal wave equation}
\indent
In this section, first we write down
transformation rules for the GSWE and recall some basic
properties of the three-term recurrence relations for the series
coefficients (not restricted to the GSWE). After this we
set the solutions for the general case
and discuss both the Whittaker-Hill and Mathieu equations from
the viewpoint of the GSWE.

In the first place, if $U(z)=U(B_{1},B_{2},B_{3}; z_{0},\omega,\eta;z)$
%
%
%
denotes one solution of the GSWE, the
transformation rules $T_{1},\ T_{2},\ T_{3}$ and
$T_{4}$ are given by \cite{eu,fisher}
\letra
\begin{eqnarray}\label{rule} \begin{array}{l}
T_{1}U(z)=z^{1+B_{1}/z_{0}}
U(C_{1},C_{2},C_{3};z_{0},\omega,\eta;z),\  \ z_{0}\neq0,
\vspace{3mm}\\
T_{2}U(z)=(z-z_{0})^{1-B_{2}-B_{1}/z_{0}}U(B_{1},D_{2},D_{3};
z_{0},\omega,\eta;z), \ \  z_{0}\neq0,
\vspace{3mm}\\
T_{3}U(z)=U(B_{1},B_{2},B_{3}; z_{0},-\omega,-\eta;z),
\ \ \forall z_{0},
\vspace{3mm}\\
T_{4}U(z)=
U(-B_{1}-B_{2}z_{0},B_{2},
B_{3}+2\eta\omega z_{0};z_{0},-\omega,
\eta;z_{0}-z),\ \  \forall z_{0}
\end{array}
\end{eqnarray}
where
\begin{eqnarray}\begin{array}{l}
C_{1}=-B_{1}-2z_{0}, \ \
C_{2}=2+B_{2}+\frac{2B_{1}}{z_{0}},\ C_{3}=B_{3}+
\left(1+\frac{B_{1}}{z_{0}}\right)
\left(B_{2}+\frac{B_{1}}{z_{0}}\right),
\vspace{.3cm}\\
D_{2}=2-B_{2}-\frac{2B_{1}}{z_{0}},\ D_{3}=B_{3}+
\frac{B_{1}}{z_{0}}\left(\frac{B_{1}}{z_{0}}
+B_{2}-1\right).
\end{array}
\end{eqnarray}
These rules can be verified by substitutions of variables.
In $T_{3}$ it is assumed that we must change
the sign of ($\eta,\omega$) only where these quantities
appear explicitly in the solution $U(z)$, keeping the expressions
for the other parameters  unchanged even if they
depend on $\eta$ and $\omega$ as in the Teukolsky
equations of the relativistic astrophysics
\cite{leaver1,bini}. For brevity, we use
only $T_{1}$ and
$T_{2}$. The rule $T_{4}$ interchanges
the position of the regular singular points ($z=z_{0}
\leftrightarrow z=0$) and may be used to get solutions
with the appropriate behavior at $z=0$.

%
On the other hand, if $b_{n}$ denotes the series coefficients
of a solution, then the three-term recurrence relations
(for solutions without phase parameter) have the form
\antiletra
\begin{eqnarray}\label{recurrence1}
\alpha_{0}b_{1}+\beta_{0}b_{0}=0,
\ \ \alpha_{n}b_{n+1}+\beta_{n}b_{n}+
\gamma_{n}b_{n-1}=0\ (n\geq1),
\end{eqnarray}
where $\alpha_{n}$, $\beta_{n}$ and $\gamma_{n}$ are
constants depending on the parameters of the
differential equation. These recurrence relations constitute
a infinite system of homogeneous linear equations which
can be written as
\begin{eqnarray}
\label{matriz}
\left(
\begin{array}{ccccccccc}
\beta_{0} & \alpha_{0} &    0       & \cdots   &     &  &   &     0 & \cdots \\
\gamma_{1}&\beta_{1}   & \alpha_{1} &  0       &     &  &   & \vdots \\
    0    &\gamma_{2} & \beta_{2}    &\alpha_{2}&     &  &   &        \\
 \vdots  &           &            &            &     &  &   &        \\
         &    &   &   &       &\gamma_{n}&\beta_{n} &\alpha_{n}&\cdots\\
         &    &   &   &       &            & & \vdots
\end{array}
\right) \left(\begin{array}{l}
b_{0}  \\
b_{1} \\
b_{2} \\
 \vdots   \\
b_{n}\\
\vdots
\end{array}
\right)=0.
\end{eqnarray}
Then, to have nontrivial solutions for $b_{n}$, the
(Hill) determinant of the above infinite tridiagonal
matrix must vanish. This is possible only if there is
some arbitrary parameter in the differential equation,
that is, in the elements of the matrix.
Equivalently, the recurrence relations lead to
characteristic equation in terms of the infinite continued
fraction \cite{leaver1}
\begin{eqnarray}\label{charcteristic1}
\beta_{0}=\frac{\alpha_{0}\gamma_{1}}{\beta_{1}-}\
\frac{\alpha_{1}\gamma_{2}}
{\beta_{2}-}\ \frac{\alpha_{2}\gamma_{3}}{\beta_{3}-}\cdots.
\end{eqnarray}
So, instead of a vanishing determinant,
we must satisfy the characteristic equation and this
also assures the series convergence
by means of some version of a Perron theorem \cite{gautschi}.

Under certain conditions, three-term recurrence
relations like (\ref{recurrence1}) give finite-series solutions
which are also called quasi-polynomial solutions, Heun
polynomials or quasi-algebraic solutions. The condition
sufficient for having terminating series is $\gamma_{n}=0$
for some $n=N=$positive integer \cite{arscott1}, in
which case the series presents $N$ terms, that is,
$0\leq n\leq N-1$ . In effect,
if $\gamma_{N}=0$  we can choose the parameters of the
equation so that $b_{N}=0$:
then, the relations (\ref{recurrence1}) imply
that $b_{n}=0$ for any $n\geq N$. In such event, in general
there are infinite series as well since the transformation
rules generate solutions with other recurrence
relations. In addition,
if $\alpha_{n}=0$ for some $n=N$, the series
begins at $n=N+1$, but in this case one may take $n= m+N+1$
and relabel the series coefficients in order to obtain
a series beginning at $m=0$.

There is another remark which will be useful
in Secs. II A and IV A. Suppose a second solution with
coefficients $c_{n}$ having the recurrence relations
\letra
\begin{eqnarray}\label{recurrence2}
\tilde{\alpha}_{0}c_{1}+\beta_{0}c_{0}=0,
\ \ \tilde{\alpha}_{n}c_{n+1}+\beta_{n}c_{n}+
\tilde{\gamma}_{n}c_{n-1}=0\ (n\geq1),
\end{eqnarray}
where $\beta_{n}$ is the same as in Eq. (\ref{recurrence1}). Then,
if
\begin{eqnarray}\label{charcteristic2}
\tilde{\alpha}_{n}\tilde{\gamma}_{n+1}=\alpha_{n}\gamma_{n+1},
\end{eqnarray}
it is obvious from Eq. (\ref{charcteristic1}) that, for infinite series,
both solutions have the same characteristic equation
provided that the sums begin at $n=0$ in both series.
%
%
%
\subsection*{A. GSWE and the Barber-Hass\'e solutions}
\indent
Now we  construct the initial set containing the
solutions ($U_{1}^{0},U_{1}^{\infty},U_{1}$).
$U_{1}^{0}$ is the Barber-Hass\'e expansion
in  series of ascending power of $z-z_{0}$ which
converges for any finite $z$ and
was originally proposed to solve an angular GSWE
in which $0\leq z \leq z_{0}$ \cite{leaver1}. $U_{1}^{\infty}$
is given by an expansion in series of irregular confluent
hypergeometric functions $\Psi(a,b;y)$ (or confluent
hypergeometric function of second kind),
converges for $\vert z\vert > \vert z_{0}\vert$ and, thus,
is suitable to solve a radial GSWE ($z\geq z_{0}$)
except at $z=z_{0}$. $U_{1}$ is
an expansion in series of regular confluent hypergeometric
functions $\Phi(a,b;y)$ (or confluent
hypergeometric function of first kind) which converges for any
value of $z$
but, in general, its behavior at infinity is not the one
predicted by the Thom\'e solutions.
From the first set of solutions we obtain three other sets by using
the transformation rules $T_{1}$ and $T_{2}$.


To obtain the Barber-Hass\'e solution $U_{1}^{0}$ we write
$U(z)=U_{1}^{0}(z)=e^{i\omega z}f(z)$. Then,
the GSWE (\ref{gswe}) leads to
\antiletra
\begin{eqnarray}\label{wilson}
z(z-z_{0})\frac{d^{2}f}{dz^{2}}+
\left[B_{1}+B_{2}z+2i\omega z(z-z_{0})\right]\frac{df}{dz}+
\left[B_{3}+i\omega B_{1}+i\omega B_{2}z-
2\omega\eta(z-z_{0})\right]f=0.
\end{eqnarray}
Now, by expanding $f(z)$ in a Frobenius series
corresponding to the exponent $0$ about $z=z_{0}$,
i. e.,
\begin{eqnarray}
f(z)=\displaystyle \sum_{n=0}^{\infty}b_{n}^{(1)}
(z-z_{0})^n
\Rightarrow U_{1}^{0}(z)=e^{i\omega z}
\displaystyle \sum_{n=0}^{\infty}b_{n}^{(1)}(z-z_{0})^n,
\end{eqnarray}
we find that $b_{n}^{(1)}$ satisfy the
recurrence relations ($b_{-1}^{(1)}=0$)
\begin{eqnarray}
&  z_{0}\left(n+B_{2}+\frac{B_{1}}{z_{0}}\right)(n+1)b_{n+1}^{(1)}+
\left[ n\left(n+B_{2}-1+2i\omega z_{0}\right)+
i\omega z_{0}\left(B_{2}+\frac{B_{1}}{z_{0}}\right)+
B_{3}\right]b_{n}^{(1)}
\nonumber \\
& +2i\omega\left(n-1+i\eta+\frac{B_{2}}{2}\right)
b_{n-1}^{(1)}=0.
\end{eqnarray}
This solution can also be obtained by
applying the rule $T_{4}$ to the Barber-Hass\'{e} expansion
given in Ref. \cite{leaver1}.
It converges for any finite $z$ provided that the characteristic
equation is fulfilled \cite{leaver1,barber}.

To find the solutions in series of confluent hypergeometric
functions, we carry out the substitutions
\letra
\begin{eqnarray}
U(z)=e^{i\omega z}F(y),\ \ y=-2i\omega z,
\end{eqnarray}
which transform the GSWE (\ref{gswe}) into
\begin{eqnarray}
\label{F}
&&(y+2i
\omega z_{0})\left[y\frac{d^2F}{dy^2}-
y\frac{dF}{dy}\right]
+B_{2}y\frac{dF}{dy}-2i\omega B_{1}\frac{dF}{dy}+
\nonumber\\
&&\hspace{1.5cm}\left[B_{3}+i\omega B_{1}+2\eta \omega z_{0}-\left(i\eta+\frac{B_{2}}{2}\right)y
\right]F=0.
\end{eqnarray}
%
%
Then, first we expand $F(y)$ in series of the irregular functions $\Psi(a,b;y)$ \cite{erdelyi1} according to
\antiletra
\begin{eqnarray}\label{F2}
\begin {array}{l}
F(y)=\displaystyle \sum_{n=0}^{\infty}c_{n}^{(1)}\Psi\left(n+i\eta+\frac{B_{2}}{2},n+B_{2};y\right)=\displaystyle \sum_{n=0}^{\infty}c_{n}^{(1)}\Psi_{n}(y)\Rightarrow
\vspace{3mm}\\
U_{1}^{\infty}(z)= e^{i\omega z}\displaystyle
\sum_{n=0}^{\infty}c_{n}^{(1)}
\Psi\left(n+i\eta+\frac{B_{2}}{2},n+B_{2};-2i\omega z\right),
\end{array}
\end{eqnarray}
where $\Psi_{n}(y)=\Psi\left(n+i\eta+B_{2}/2,n+B_{2};y\right)$.
These $\Psi_{n}(y)$ satisfy
\begin{eqnarray}\begin{array}{l}\label{F3}
y\frac{d^2\Psi_{n}(y)}{dy^2}-y\frac{d\Psi_{n}(y)}{d
y}=-(n+B_{2})\frac{d\Psi_{n}(y)}{dy}+
\left(n+i\eta+\frac{B_{2}}{2}\right)\Psi_{n}(y),\vspace{3mm}\\
\frac{d\Psi_{n}(y)}{dy}=-\left(n+i\eta+
\frac{B_{2}}{2}\right)\Psi_{n+1}(y),\vspace{3mm}\\
y\frac{d\Psi_{n}(y)}{dy}=(1-n-B_{2}+y)\Psi_{n}(y)- \Psi_{n-1}(y),
\vspace{3mm}\\
\left(n+i\eta+\frac{B_{2}}{2}\right)y \Psi_{n+1}(y)+(1-n-B_{2}+y)\Psi_{n}(y)-\Psi_{n-1}(y)=0,
\end{array}\end{eqnarray}
where the first relation is the confluent hypergeometric equation
for $\Psi_{n}(y)$ and the
others follow from the properties of the
irregular confluent hypergeometric functions \cite{erdelyi1}.
Thus, inserting the previous expression for
$F(y)$ into Eq. (\ref{F})
and using Eqs. (\ref{F3}), we obtain
\begin{eqnarray}\label{irregular}
\displaystyle \sum_{n=1}^{\infty}nc_{n}^{(1)}\Psi_{n-1}(y)+
\displaystyle \sum_{n=0}^{\infty}c_{n}^{(1)}\left[\beta_{n}^{(1)}\Psi_{n}(y)+2i\omega z_{0}\left(n+B_{2}+\frac{B_{1}}{z_{0}}\right) \left(n+i\eta+\frac{B_{2}}{2}\right)
\Psi_{n+1}(y)\right]=0,
\end{eqnarray}
where $\beta_{n}^{(1)}$ is the coefficient of $b_{n}^{(1)}$
in the recurrence relation for the solution $U_{1}^{0}$, namely,
\begin{eqnarray*}
\beta_{n}^{(1)}=n\left(n+B_{2}-1+2i\omega z_{0}\right)+
i\omega z_{0}\left(B_{2}+\frac{B_{1}}{z_{0}}\right)+
B_{3}.
\end{eqnarray*}
Putting $m=n-1$ and $m=n+1$ in the first and last terms
of Eq. (\ref{irregular}), respectively, we find
\begin{eqnarray*}
\begin{array}{c}
\left[ c_{1}^{(1)}+\beta_{0}^{(1)}c_{0}^{(1)}\right] \Psi_{0}(y)+
\vspace{3mm}\\
\displaystyle\sum_{m=1}^{\infty}\left[
(m+1)c_{m+1}^{(1)}+
\beta_{m}^{(1)}c_{m}^{(1)}+
2i\omega z_{0}
\left(m-1+B_{2}+\frac{B_{1}}{z_{0}}\right)
\left(m-1+i\eta+\frac{B_{2}}{2}\right)
c_{m-1}^{(1)}
\right]\Psi_{m}(y)=0.
\end{array}
\end{eqnarray*}
By equating to zero the coefficients of each
independent $\Psi_{n}(y)$ we find the three-term
recurrence relations ($c_{-1}^{(1)}=0$)
\begin{eqnarray}\label{c1}
(n+1)c_{n+1}^{(1)}+\beta_{n}^{(1)}c_{n}^{(1)}+
2i\omega z_{0}\left(n-1+B_{2}+\frac{B_{1}}{z_{0}}\right) \left(n-1+i\eta+\frac{B_{2}}{2}\right)
c_{n-1}^{(1)}=0,\ \ n\geq 0.
\end{eqnarray}

From the behavior of $\Psi(a,b;y)$, when
$b\rightarrow \infty$ while $b-a$ and $y$ remain
bounded \cite{erdelyi1}, we get
\begin{eqnarray*}
\lim_{n\rightarrow \infty}\frac{\Psi_{n+1}(y)}{\Psi_{n}(y)}=\frac{1}{2i\omega z}.
\end{eqnarray*}
Besides this, using a Perron-Kreuser
theorem \cite{gautschi} we find that the minimal solution
of the recurrence relations (\ref{c1}) satisfies
\begin{eqnarray*}
\lim_{n\rightarrow \infty}\frac{c_{n+1}^{(1)}}{c_{n}^{(1)}}=
\left\{
\begin{array}{lr}
-2i\omega z_{0}, \ \ &\mbox{if}\ z_{0}\neq 0; \vspace{3mm}\\
-2i\omega B_{1}/n, \ \ &\mbox{if}\ z_{0}=0.
\end{array}\right.
\end{eqnarray*}
Then, combining these two limits, we have
\begin{eqnarray}
\lim_{n\rightarrow \infty}\frac{c_{n+1}^{(1)}\Psi_{n+1}(y)}{c_{n}^{(1)}\Psi_{n}(y)}=
\left\{
\begin{array}{lr}
-z_{0}/z, \ \ &\mbox{if}\ z_{0}\neq 0; \vspace{3mm}\\
-B_{1}/(nz), \ \ &\mbox{if}\ z_{0}=0.
\end{array}\right.
\end{eqnarray}
Therefore, by the ratio test, the series in $U_{1}^{\infty}(z)$
converges for any $|z|>|z_{0}|$ on the condition that
the characteristic equation is satisfied.
The behavior of $U_{i}^{\infty}(z)$ when $z\rightarrow \infty$
results from the relation \cite{erdelyi1}
\begin{eqnarray}
\label{asymptotic}
\lim_{y\rightarrow \infty}\Psi(a,b;y)\sim y^{-a}[1+O(|y|^{-1})], \ \ -\frac{3\pi}{2}< \arg{y}<-\frac{3\pi}{2},
\end{eqnarray}
which implies
\begin{eqnarray*}
\lim_{z\rightarrow  \infty}U_{1}^{\infty}(z)
\sim e^{i\omega z}z^{- i\eta-(B_{2}/2)},
 \ \ -\frac{3\pi}{2}<\arg{(-2 i \omega z)}<\frac{3\pi}{2},
\end{eqnarray*}
as in a normal Thom\'e solution.

%
%
Now we expand $F(y)$ in series of regular confluent hypergeometric functions $\Phi(a,b;y)$.
If we define the function $\hat{\Phi}_{n}(y)$ as
\begin{eqnarray}
\hat{\Phi}_{n}(y)=
\frac{(-1)^{n}}{\Gamma(n+B_{2})}
\Phi\left(n+i\eta+\frac{B_{2}}{2},n+B_{2};y\right),
y=-2i\omega z
\end{eqnarray}
and use some difference and differential properties
of $\Phi(a,b;y)$  \cite{abramowitz}, we find
that $\hat{\Phi}_{n}(z)$ also satisfy the relations (\ref{F3}).
Therefore, substituting  $\hat{\Phi}_{n}(y)$  for $\Psi_{n}(y)$ in
Eq. (\ref{F2}) we get the solution
\begin{eqnarray}
U_{1}(z)=e^{i\omega z}
\sum_{n=0}^{\infty} c_{n}^{(1)}\hat{\Phi}_{n}(y)=e^{i\omega z}
\sum_{n=0}^{\infty} c_{n}^{(1)}
\hat{\Phi}\left(n+i\eta+\frac{B_{2}}{2},n+B_{2};-2i\omega z\right),
\end{eqnarray}
where the coefficients $c_{n}^{(1)}$ are the same
which appear in the solution $U_{1}^{\infty}(z)$.
In order to show that the series in $U_{1}(z)$
converges for any $z$, we use the  behavior of $\Phi(a,b;y)$
when $b\rightarrow \infty$ while $b-a$ and $y$ remain
bounded \cite{erdelyi1}, namely,
\begin{eqnarray*}
\lim_{b\rightarrow \infty}\Phi(a,b;y)=e^{y}[1+O(|b|^{-1})],
\ \ (b-a=\mbox{finite})
\end{eqnarray*}
which leads to
\begin{eqnarray*}
\lim_{n\rightarrow \infty}\frac{\hat{\Phi}_{n+1}(y)}
{\hat{\Phi}_{n}(y)}=-\frac{1}{n+B_{2}}.
\end{eqnarray*}
Thus, using again the minimal solution of the recurrence
relations for $c_{n}^{(1)}$ we get
\begin{eqnarray*}
\lim_{n\rightarrow \infty}\frac{c_{n+1}^{(1)}\hat{\Phi}_{n+1}(y)}
{c_{n}^{(1)}\hat{\Phi}_{n}(y)}=
\left\{
\begin{array}{lr}
\frac{2i\omega z_{0}}{n+B_{2}}\rightarrow 0, \ \ &\mbox{if}\ z_{0}\neq 0; \vspace{3mm}\\
\frac{2i\omega B_{1}}{n(n+B_{2})}\rightarrow 0, \ \ &\mbox{if}\ z_{0}=0.
\end{array}\right.
\end{eqnarray*}
Therefore, the ratio test implies that the series in the
solution $U_{1}(z)$ converges for any value of $z$.
However, the behavior of $U_{1}(z)$ when
$z\rightarrow\infty$ in general does not coincide with
the one inferred from the Thom\'e solutions, since
we have \cite{abramowitz}
\begin{eqnarray}
\label{regular}
\lim_{y\rightarrow \infty}\Phi(a,b;y)=
\left\{
\begin{array}{l}
\frac{\Gamma(b)}{\Gamma(a)}e^{y}y^{a-b}[1+O(|y|^{-1})], \ \ \ \ \ \ (\Re{y}>0)
\vspace{3mm}\\
\frac{\Gamma(b)}{\Gamma(b-a)}(-y)^{-a}[1+O(|y|^{-1})], \ \  (\Re{y}<0).
\end{array}
\right.
\end{eqnarray}
Thus, only if $\Re{y}<0$ the asymptotic behavior of
$U_{1}$ has the same form as $U_{1}^{\infty}$. If
$\Re{y}=0$, the limit of $\Phi(a,b;y)$ is a combination
of the two expressions given on the right-hand side of
the previous expression \cite{erdelyi1}. On the other hand,
although the function $\Phi(a,b;y)$
is not defined if $b=0,-1,-2,\cdots$, the function
$\hat{\Phi}(a,b;y)$ is defined since \cite{erdelyi1}
\begin{eqnarray}
\displaystyle \lim_{b\rightarrow 1-m}\frac{\Phi(a,b;y)}{\Gamma{(b)}}=
\frac{(a)_{m}}{m!}\ y^{m}\ \Phi(a+m,1+m;y), \ \ \ m=1,2,3,\cdots,
\end{eqnarray}
$(a)_{m}=a(a+1)\cdots (a+m-1)$
for $m\geq1$. Therefore, the expansion $U_{1}(z)$ is
defined even if $B_{2}$ is zero or a negative integer.

In the following we collect the
three solutions in the first set of solutions, and this
gives three other sets by the use of
the transformation rules $T_{1}$ and $T_{2}$ according to
\begin{eqnarray}\label{rules}
\left(U_{1}^{0},U_{1}^{\infty},
U_{1}\right)
\stackrel{T_{1}}{\longleftrightarrow}
\left(U_{2}^{0},U_{2}^{\infty},U_{2}\right)
\stackrel{T_{2}}{\longleftrightarrow}
\left(U_{3}^{0},U_{3}^{\infty},U_{3}\right)
\stackrel{T_{1}}{\longleftrightarrow}
\left(U_{4}^{0},U_{4}^{\infty},U_{4}\right)
\stackrel{T_{2}}{\longleftrightarrow}
\left(U_{1}^{0},U_{1}^{\infty},U_{1}\right).
\end{eqnarray}
%
%
%
{\it First set}. This set admits both the Leaver and Ince limits.
If $B_{2}+B_{1}/z_{0}=1$, it is equal to the fourth set.
\letra
\begin{eqnarray}
\label{frobenius1a} \begin{array}{l}
U_{1}^{0}(z)=e^{i\omega z}\displaystyle \sum_{n=0}^{\infty}b_{n}^{(1)}
(z-z_{0})^{n},
\vspace{3mm}\\
U_{1}^{\infty}(z)=e^{i\omega z}\displaystyle \sum_{n=0}^{\infty}c_{n}^{(1)}\Psi\left(n+i\eta+\frac{B_{2}}{2},n+B_{2};-2i\omega z\right),
\vspace{3mm}\\
U_{1}(z)=e^{i\omega z}\displaystyle
\sum_{n=0}^{\infty} c_{n}^{(1)}
\hat{\Phi}\left(n+i\eta+\frac{B_{2}}{2},n+B_{2};-2i\omega z\right).
\end{array}\end{eqnarray}
Setting $b_{-1}^{(1)}=0$ and $c_{-1}^{(1)}=0$,
the three-term recurrence relations for the
coefficients are given by
\begin{eqnarray}\label{fronenius1c}
&  z_{0}\left(n+B_{2}+\frac{B_{1}}{z_{0}}\right)(n+1)b_{n+1}^{(1)}+
\left[ n\left(n+B_{2}-1+2i\omega z_{0}\right)+
i\omega z_{0}\left(B_{2}+\frac{B_{1}}{z_{0}}\right)+
B_{3}\right]b_{n}^{(1)}
\nonumber \\
& +2i\omega\left(n-1+i\eta+\frac{B_{2}}{2}\right)
b_{n-1}^{(1)}=0,
\end{eqnarray}
and
\begin{eqnarray}
&&(n+1)c_{n+1}^{(1)}+
\left[ n\left(n+B_{2}-1+2i\omega z_{0}\right)+
i\omega z_{0}\left(B_{2}+\frac{B_{1}}{z_{0}}\right)+
B_{3}\right]c_{n}^{(1)}+
\nonumber\vspace{3mm}\\
&&2i\omega z_{0}\left(n-1+B_{2}+\frac{B_{1}}{z_{0}}\right) \left(n-1+i\eta+\frac{B_{2}}{2}\right)c_{n-1}^{(1)}=0.
\end{eqnarray}
The coefficients $b_{n}^{(1)}$ and $c_{n}^{(1)}$
are connected by
\antiletra
\begin{eqnarray}\label{connect1}
c_{n}^{(1)}=\left( z_{0}\right) ^{n}
\Gamma\left(n+B_{2}+\frac{B_{1}}{z_{0}}\right)
b_{n}^{(1)}, \ \mbox{if} \ B_{2}+\frac{B_{1}}{z_{0}}\neq 0,-1,-2,\cdots
\end{eqnarray}
apart from a multiplicative constant independent
of $n$. The restrictions on $B_{2}+(B_{1}/z_{0})$ assure
that the sum in the three solutions
begins at $n=0$ and, thus, they have the same
characteristic equation for the reason given
in the paragraph containing Eqs. (\ref{recurrence2})
and (\ref{charcteristic2}). In addition, we have finite-series
solutions if $i\eta+(B_{2}/2)$ is zero or negative integer.
%
%
%
%

{\it Second set}. These solutions admit the Ince
limit but not the Leaver limit.
If $B_{2}+B_{1}/z_{0}=1$, the second and the third sets are
equal to one another.
\letra
\begin{eqnarray}\begin{array}{l}
U_{2}^{0}(z)=e^{i\omega z}z^{1+\frac{B_{1}}{z_{0}}}
\displaystyle \sum_{n=0}^{\infty}b_{n}^{(2)}(z-z_{0})^{n},
\vspace{3mm}\\
U_{2}^{\infty}(z)=e^{i\omega z}z^{1+\frac{B_{1}}{z_{0}}}
\displaystyle\sum_{n=0}^{\infty}c_{n}^{(2)}
\Psi\left(n+1+i\eta+\frac{B_{1}}{z_{0}}+\frac{B_{2}}{2},
n+2+B_{2}+\frac{2B_{1}}{z_{0}};-2i\omega z\right),
\vspace{3mm}\\
U_{2}(z)=e^{i\omega z}z^{1+\frac{B_{1}}{z_{0}}}
\displaystyle\sum_{n=0}^{\infty}
 c_{n}^{(2)}
\hat{\Phi}\left(
n+1+i\eta+\frac{B_{1}}{z_{0}}+\frac{B_{2}}{2},
n+2+B_{2}+\frac{2B_{1}}{z_{0}};-2i\omega z\right),
\end{array}\end{eqnarray}
where the recurrence relations are
\begin{eqnarray}
&z_{0}\left(n+B_{2}+\frac{B_{1}}{z_{0}}\right)(n+1)b_{n+1}^{(2)}+
\nonumber\\
&\left[n\left(n+1+2i\omega z_{0}+B_{2}
+\frac{2B_{1}}{z_{0}}\right)+
i\omega z_{0}\left(B_{2}+\frac{B_{1}}{z_{0}}\right)+
\left(1+\frac{B_{1}}{z_{0}}\right)\left(B_{2}+
\frac{B_{1}}{z_{0}}\right)+B_{3}\right]b_{n}^{(2)}
\nonumber\\
&+2i\omega \left(n+i\eta+\frac{B_{1}}{z_{0}}+\frac{B_{2}}{2}\right)
b_{n-1}^{(2)}=0,
\end{eqnarray}
and
\begin{eqnarray}
&(n+1)c_{n+1}^{(2)}+
\nonumber\\
&\left[n\left(n+1+2i\omega z_{0}+B_{2}
+\frac{2B_{1}}{z_{0}}\right)+
i\omega z_{0}\left(B_{2}+\frac{B_{1}}{z_{0}}\right)+
\left(1+\frac{B_{1}}{z_{0}}\right)\left(B_{2}+
\frac{B_{1}}{z_{0}}\right)+B_{3}\right]c_{n}^{(2)}
\nonumber\\
&+2i\omega\left(n-1+B_{2}+\frac{B_{1}}{z_{0}}\right)
\left(n+i\eta+\frac{B_{1}}{z_{0}}+\frac{B_{2}}{2}\right)
c_{n-1}^{(2)}=0.
\end{eqnarray}
The coefficients are connected by
\antiletra
\begin{eqnarray}\label{connect2}
c_{n}^{(2)}=\left( z_{0}\right) ^{n}\Gamma\left(n+B_{2}+
\frac{B_{1}}{z_{0}}\right)b_{n}^{(2)}, \ \mbox{if} \
B_{2}+\frac{B_{1}}{z_{0}}\neq -0,-1,-2,\cdots.
\end{eqnarray}
%
%

{\it Third set}. These solutions admit both the Leaver and Ince limits.
\letra
\begin{eqnarray}\label{frobenius3} \begin{array}{l}
U_{3}^{0}(z)=e^{i\omega z}z^{1+\frac{B_{1}}{z_{0}}}
(z-z_{0})^{1-B_{2}-\frac{B_{1}}{z_{0}}}
\displaystyle \sum_{n=0}^{\infty}b_{n}^{(3)}(z-z_{0})^{n},
\vspace{3mm}\\
U_{3}^{\infty}(z)=e^{i\omega z}z^{1+\frac{B_{1}}{z_{0}}}
(z-z_{0})^{1-B_{2}-\frac{B_{1}}{z_{0}}}
\displaystyle \sum_{n=0}^{\infty}
c_{n}^{(3)}\Psi\left(n+2+i\eta-\frac{B_{2}}{2},
n+4-B_{2};-2i\omega z\right),
\vspace{3mm}\\
U_{3}(z)=e^{i\omega z}z^{1+\frac{B_{1}}{z_{0}}}
(z-z_{0})^{1-B_{2}-\frac{B_{1}}{z_{0}}}
\displaystyle\sum_{n=0}^{\infty}c_{n}^{(3)}
\hat{\Phi}\left(n+2+i\eta-\frac{B_{2}}{2},n+4-B_{2};-2i\omega z\right),
\end{array}
\end{eqnarray}
where the recurrence relations are given by
\begin{eqnarray}
&z_{0}\left(n+2-B_{2}-\frac{B_{1}}{z_{0}}\right)(n+1)b_{n+1}^{(3)}
\nonumber\\
&+\left[n\left(n+3+2i\omega z_{0}-B_{2}\right)+
i\omega z_{0} \left(2-B_{2}-\frac{B_{1}}{z_{0}}\right)+2-B_{2}+B_{3}
\right]b_{n}^{(3)}
\nonumber \\
&+2i\omega \left(n+1+i\eta-\frac{B_{2}}{2}\right)b_{n-1}^{(3)}=0
\end{eqnarray}
and
\begin{eqnarray}
&(n+1)c_{n+1}^{(3)}
+\left[n\left(n+3+2i\omega z_{0}-B_{2}\right)+
i\omega z_{0}
 \left(2-B_{2}-\frac{B_{1}}{z_{0}}\right)+2-B_{2}+B_{3}
\right]c_{n}^{(3)}+
\nonumber \\
&2i\omega\left(n+1-B_{2}-\frac{B_{1}}{z_{0}}\right)
\left(n+1+i\eta-\frac{B_{2}}{2}\right)c_{n-1}^{(3)}=0.
\end{eqnarray}
In this case, we find
\antiletra
\begin{eqnarray}\label{connect3}
c_{n}^{(3)}=(z_{0})^{n}\Gamma\left( n+2-B_{2}-\frac{B_{1}}{z_{0}}\right)
b_{n}^{(3)}, \ \mbox{if}\  B_{2}+\frac{B_{1}}{z_{0}}\neq 2,3,4,\cdots.
\end{eqnarray}
%

%
%
%
%
{\it Fourth set}.
This set admits the Ince limit but not the Leaver limit.
\letra
\begin{eqnarray}\begin{array}{l}
U_{4}^{0}(z)=e^{i\omega
 z}(z-z_{0})^{1-B_{2}-\frac{B_{1}}{z_{0}}}
\displaystyle \sum_{n=0}^{\infty}b_{n}^{(4)}(z-z_{0})^{n},
\vspace{3mm}\\
U_{4}^{\infty}(z)=e^{i\omega z}(z-z_{0})^{1-B_{2}-\frac{B_{1}}{z_{0}}}\displaystyle \sum_{n=0}^{\infty}c_{n}^{(4)}
\Psi\left(n+1+i\eta-\frac{B_{1}}{z_{0}}-\frac{B_{2}}{2},n+2-B_{2}-
\frac{2B_{1}}{z_{0}};-2i\omega z\right),
\vspace{3mm}\\
U_{4}(z)=
e^{i\omega z}(z-z_{0})^{1-B_{2}-\frac{B_{1}}{z_{0}}}
\displaystyle \sum_{n=0}^{\infty}
 c_{n}^{(4)}
\hat{\Phi}\left(n+1+i\eta-\frac{B_{1}}{z_{0}}-\frac{B_{2}}{2},n+2-B_{2}-
\frac{2B_{1}}{z_{0}};-2i\omega z\right),
\end{array}
\end{eqnarray}
with the recurrence relations
\begin{eqnarray}
&z_{0}\left(n+2-B_{2}-\frac{B_{1}}{z_{0}}\right)(n+1)b_{n+1}^{(4)}+
\nonumber\\
&\left[n\left(n+1+2i\omega z_{0}-B_{2}-
\frac{2B_{1}}{z_{0}}\right)+i\omega z_{0}
\left(2-B_{2}-\frac{B_{1}}{z_{0}}\right)+
\frac{B_{1}}{z_{0}}\left(B_{2}+\frac{B_{1}}{z_{0}}-1\right)
+B_{3}\right]
\nonumber \\
&\times b_{n}^{(4)}+2i\omega \left(n+i\eta-\frac{B_{1}}{z_{0}}-\frac{B_{2}}{2}\right)
b_{n-1}^{(4)}=0,
\end{eqnarray}
and
\begin{eqnarray}
&(n+1)c_{n+1}^{(4)}+
\nonumber\\
&\left[n\left(n+1+2i\omega z_{0}-B_{2}-
\frac{2B_{1}}{z_{0}}\right)+i\omega z_{0}
\left(2-B_{2}-\frac{B_{1}}{z_{0}}\right)+
\frac{B_{1}}{z_{0}}\left(B_{2}+\frac{B_{1}}{z_{0}}-1\right)
+B_{3}\right]c_{n}^{(4)}+
\nonumber \\
&
2i\omega z_{0}\left(n+1-B_{2}-\frac{B_{1}}{z_{0}}\right)
\left(n+i\eta-\frac{B_{1}}{z_{0}}-\frac{B_{2}}{2}\right)
c_{n-1}^{(4)}=0.
\end{eqnarray}
This time the connecting formula is
\antiletra
\begin{eqnarray}\label{connect4}
c_{n}^{(4)}=(z_{0})^{n}
\Gamma\left( n+2-B_{2}-\frac{B_{1}}{z_{0}}\right) b_{n}^{(4)},
\ \mbox{if}\  B_{2}+\frac{B_{1}}{z_{0}}\neq 2,3,4,\cdot.
\end{eqnarray}

As the rule $T_{4}$ changes
the position of the regular singular points, it may be used
to get solutions with the suitable behavior at $z=0$.
For these new sets of solutions, the $U_{i}^{0}$ ($i=5,6,7,8$)
again converge for any finite value of $z$, the $U_{i}^{\infty}$
converge for $|z-z_{0}|>$ $|z_{0}|$, whereas the $U_{i}$
converge for all values of $z$.

The previous convergence properties are
transferred to the limiting and particular cases of the GSWE.
The utility of the convergence of the solutions $U_{i}$
over the entire complex plane requires further investigation.
In fact, other solutions converging for any $z$
(also giving by series of regular confluent
hypergeometric functions) have already been found
in Sec. VII of Ref. \cite{leaver1}. In any case, in
physical problems we have also to consider boundary and
regularity conditions, in addition to convergence.

%
%
\subsection*{B. Whittaker-Hill  equation as a GSWE and Arscott's solutions}
\indent
By inserting
$z=\cos^{2}(\kappa u)$ and $W(u)=U(z)$
into the WHE (\ref{whe}), we obtain the GSWE
\begin{eqnarray}\label{compare}
z(z-1)\frac{d^{2}U}{dz^{2}}+\left(-\frac{1}{2}+z\right)
\frac{dU}{dz}+
\left[\frac{1}{4}(p+1)\xi-\frac{\vartheta}{4}+
\frac{(p+1)\xi}{2}(z-1)-\frac{\xi^2}{4}z(z-1)\right]U=0.
\end{eqnarray}
From this and Eq. (\ref{gswe}), we find that  the solutions
$W(u)$ for the WHE (\ref{whe}) may be written in terms of
the solutions $U(z)$ of the GSWE (\ref{gswe}) as follows
\begin{eqnarray}\label{wheasgswe}
\left.
\begin{array}{l}
W(u)=U(z), \ \  \ z=\cos^{2}(\kappa u)\   (\kappa=1,i);
\vspace{3mm}\\
z_{0}=1,\ B_{1}=-\frac{1}{2}, \ B_{2}=1, \
B_{3}=\frac{1}{4}[(p+1)\xi-\vartheta], \ i\omega=\frac{\xi}{2},
\ i\eta=\frac{p+1}{2}
\end{array}\right\}
\begin{array}{l}
\mbox{WHE  as}\vspace{3mm}\\
\mbox{GSWE.}
\end{array}
\end{eqnarray}
To obtain the solutions for the WHE in terms of $z$
it is sufficient to replace the parameters given above
into the solutions $U(z)$ of the GSWE, but to
get the periodicity and parity properties of the solutions
we also use the transformation $z=\cos^{2}(\kappa u)$.

\indent
{\it First set}. Even solutions, period $\pi$ if $\kappa=1$
and $u=$real (if $\kappa=1$ $W_{1}^{\infty}$ does
not converge).
\letra
\begin{eqnarray}\begin{array}{ll}
W_{1}^{0}(u)=e^{\frac{\xi}{2}\cos^2(\kappa u)}
\displaystyle \sum_{n=0}^{\infty}\frac{(-1)^{n}}{\Gamma[n+(1/2)]}c_{n}^{(1)}
\sin^{2n}(\kappa u),& |\cos(\kappa{u})|<\infty,
\vspace{3mm}\\
W_{1}^{\infty}(u)=e^{\frac{\xi}{2}\cos^2(\kappa u)}\displaystyle \sum_{n=0}^{\infty}
c_{n}^{(1)}
\Psi\left(n+1+\frac{p}{2},n+1;-\xi \cos^{2}(\kappa u)\right),
&|\cos(\kappa u)|>1,
\vspace{3mm}\\
W_{1}(u)=e^{\frac{\xi}{2}\cos^2(\kappa u)}\displaystyle \sum_{n=0}^{\infty}c_{n}^{(1)}
\hat{\Phi}\left(n+1+\frac{p}{2},n+1;-\xi \cos^{2}(\kappa u)\right),
&\forall u
\end{array}
\end{eqnarray}
where the recurrence relations for $c_{n}^{(1)}$ are
\begin{eqnarray}
(n+1)c_{n+1}^{(1)}+
\left[n\left(n+\xi\right)+\frac{2\xi+p\xi-\vartheta}{4}
\right]c_{n}^{(1)}
+\xi \left(n-\frac{1}{2}\right) \left(n+\frac{p}{2}\right)c_{n-1}^{(1)}=0.
\end{eqnarray}
\indent
{\it Second set}. Even solutions with period $2\pi$ if $\kappa=1$.
\antiletra\letra
\begin{eqnarray}\begin{array}{ll}
W_{2}^{0}(u)=\cos(\kappa u)\ e^{\frac{\xi}{2}\cos^2(\kappa u)}
\displaystyle \sum_{n=0}^{\infty}\frac{(-1)^{n}}{\Gamma[n+(1/2)]}c_{n}^{(2)}
\sin^{2n}(\kappa u),& |\cos(\kappa{u})|<\infty,
\vspace{3mm}\\
W_{2}^{\infty}(u)=\cos(\kappa u)\ e^{\frac{\xi}{2}\cos^2(\kappa u)}
\displaystyle \sum_{n=0}^{\infty}
c_{n}^{(2)}
\Psi\left(n+\frac{3}{2}+\frac{p}{2},n+2;-\xi\cos^{2}(\kappa u)\right),&|\cos(\kappa{u})|>1,
\vspace{3mm}\\
W_{2}(u)=\cos(\kappa u)\ e^{\frac{\xi}{2}\cos^2(\kappa u)}
\displaystyle \sum_{n=0}^{\infty}
c_{n}^{(2)}
\hat{\Phi}\left(n+\frac{3}{2}+\frac{p}{2},n+2;-\xi\cos^{2}(\kappa u)\right),&\forall u
\end{array}\end{eqnarray}
with the recurrence relations
\begin{eqnarray}
(n+1)c_{n}^{(2)}+
\left[n\left(n+1+\xi\right)+\frac{1+2\xi+p\xi-\vartheta}{4}
\right]c_{n}^{(2)}
+\xi\left(n-\frac{1}{2}\right)
\left(n+\frac{p}{2}+\frac{1}{2}\right)c_{n-1}^{(2)}=0.
\end{eqnarray}
\indent
{\it Third set}. Odd solutions, period $\pi$ if $\kappa=1$.
\antiletra\letra
\begin{eqnarray}\begin{array}{ll}
W_{3}^{0}(u)=\sin(2\kappa u)\ e^{\frac{\xi}{2}\cos^2(\kappa u)}
\displaystyle \sum_{n=0}^{\infty}\frac{(-1)^{n}}{\Gamma[n+(3/2)]}c_{n}^{(3)}
\sin^{2n}(\kappa u),& |\cos(\kappa{u})|<\infty,
\vspace{3mm}\\
W_{3}^{\infty}(u)=\sin(2\kappa u)\ e^{\frac{\xi}{2}\cos^2(\kappa u)}
\displaystyle \sum_{n=0}^{\infty}
c_{n}^{(3)}
\Psi\left(n+2+\frac{p}{2},n+3;-\xi
\cos^{2}(\kappa u)\right),& |\cos(\kappa{u})|>1,
\vspace{3mm}\\
W_{3}(u)=\sin(2\kappa u)\ e^{\frac{\xi}{2}\cos^2(\kappa u)}
\displaystyle \sum_{n=0}^{\infty}c_{n}^{(3)}
\hat{\Phi}\left(n+2+\frac{p}{2},n+3;-\xi
\cos^{2}(\kappa u)\right),&\forall u
\end{array}\end{eqnarray}
having the recurrence relations
\begin{eqnarray}
 (n+1) c_{n+1}^{(3)}+
\left[n\left(n+2+\xi\right)+1
+\xi+\frac{p\xi-\vartheta}{4}\right]c_{n}^{(3)}
+\xi\left(n+\frac{1}{2}\right)\left(n+1+\frac{p}{2}\right)
c_{n-1}^{(3)}=0.
\end{eqnarray}
\indent
{\it Fourth set}. Odd solutions with period $2\pi$ if $\kappa=1$.
\antiletra\letra
\begin{eqnarray}\begin{array}{ll}
W_{4}^{0}(u)=\sin(\kappa u)\ e^{\frac{\xi}{2}\cos^2(\kappa u)}
\displaystyle \sum_{n=0}^{\infty}\frac{(-1)^{n}}{\Gamma[n+(3/2)]}c_{n}^{(4)}
\sin^{2n}(\kappa u),& |\cos(\kappa{u})|<\infty,
\vspace{3mm}\\
W_{4}^{\infty}(u)=\sin(\kappa u)\ e^{\frac{\xi}{2}\cos^2(\kappa u)}
\displaystyle \sum_{n=0}^{\infty}
c_{n}^{(4)}
\Psi\left(n+\frac{3}{2}+\frac{p}{2},n+2;-\xi \cos^{2}(\kappa u)\right),& |\cos(\kappa{u})|>1,
\vspace{3mm}\\
W_{4}(u)=\sin(\kappa u)\ e^{\frac{\xi}{2}\cos^2(\kappa u)}
\displaystyle \sum_{n=0}^{\infty}c_{n}^{(4)}
\hat{\Phi}\left(n+\frac{3}{2}+\frac{p}{2},n+2;-\xi \cos^{2}(\kappa u)\right),&\forall u
\end{array}\end{eqnarray}
where the recurrence relations are
\begin{eqnarray}
(n+1)c_{n+1}^{(4)}+
\left[ n\left(n+1+\xi\right)+\xi+\frac{1+p\xi-\vartheta}{4}
\right]c_{n}^{(4)}+\xi\left(n+\frac{1}{2}\right)
 \left(n+\frac{1}{2}+\frac{p}{2}\right)
c_{n-1}^{(4)}=0.
\end{eqnarray}

The solutions corresponding to the opposite choice for the
signs of $\omega$ and $\eta$ in Eqs. (\ref{wheasgswe})
are obtained by the substitutions
\antiletra\letra
\begin{eqnarray}
\xi\rightarrow -\xi,\ p\rightarrow -p-2,
\end{eqnarray}
which take the place of the rule $T_{3}$. The operation
equivalent to the rule $T_{4}$ is
\begin{eqnarray}
u\rightarrow u+\frac{\pi}{2\kappa},\ \
\xi\rightarrow -\xi.
\end{eqnarray}
Similarly, the composition $T_{3}T_{4}$ corresponds to
the symmetry
\begin{eqnarray}
u\rightarrow u+\frac{\pi}{2\kappa},\ \
p\rightarrow -p-2,
\end{eqnarray}
of the WHE. In fact, if we apply this last operation
to the preceding sets of solutions, we find the solutions
$W_{i}^{0}$ obtained by Arscott \cite{arscott},
each one accompanied
by two expansions in series of hypergeometric functions.

Notice that the use of the transformations $T_{1}$ and  $T_{2}$
have been essential to generate a subgroup containing four sets of
solutions which are even or odd and have period $\pi$ or $2\pi$.
The solutions for WHE obtained from the one for DCHE
do not possess these properties of parity and periodicity
(see Sec. III B).
%
%
\subsection*{C. Mathieu equation as a GSWE}
\indent
Although the Whittaker-Ince limit of the preceding solutions for
Whittaker-Hill equation gives solutions for the Mathieu equation,
such solutions are also particular cases of the solutions
for the Whittaker-Ince limit of the GSWE and, for this reason, will be
discussed in Sec. IV B. Here we regard the solutions
$W(u)$ for the Mathieu equation (\ref{mathieu})
as a particular case (instead of a limiting case) of the
solutions for the GSWE. Such solutions $W(u)$ may
be obtained by
\antiletra
\begin{eqnarray}\label{mathieu1}
\left.
\begin{array}{l}
W(u)=U(z), \ z=\cos^{2}(\frac{\sigma u}{2}),\   \ (\sigma=1,i),
\vspace{3mm}\\
z_{0}=1,\ B_{1}=-\frac{1}{2}, \ B_{2}=1, \
B_{3}=2k^2-a, \ \omega=\pm4 k,\ \
\eta=0
\end{array}\right\}
\begin{array}{l}
\mbox{Mathieu equation}\vspace{3mm}\\
\mbox{as GSWE,}
\end{array}
\end{eqnarray}
where $U(z)$ are the solutions for GSWE given
in Sec. II A, or, equivalently, by writing
\begin{eqnarray}\label{mathieu2}
 \kappa=\sigma/2, \ p=-1, \ \vartheta=4a-8k^2, \
\xi=\pm8i k\ \ (\mbox{Mathieu equation as WHE})
\end{eqnarray}
into the solutions $W(u)$ for the WHE, given in Sec. II B.
These solutions are also even or odd but their period are twice
the period of the solutions corresponding to the WHE
(if $\sigma=1$), that is, one has period $2\pi$ and $4\pi$.
As usual, there is no finite-series solutions.

{\it First set}. Even solutions with period $2\pi$ if $\sigma=1$.
\letra
\begin{eqnarray}\begin{array}{ll}
W_{1}^{0}(u)=e^{\pm4ik\cos^2\left(\sigma u/2\right)}
\displaystyle \sum_{n=0}^{\infty}
\frac{(-1)^{n}}{\Gamma[n+(1/2)]}c_{n}^{(1)}
\sin^{2n}\frac{\sigma u}{2},& \left|\cos\frac{\sigma{u}}{2}\right|<\infty,
\vspace{3mm}\\
W_{1}^{\infty}(u)=e^{\pm4ik\cos^2\left(\sigma u/2\right)}\displaystyle \sum_{n=0}^{\infty}
c_{n}^{(1)}
\Psi\left(n+\frac{1}{2},n+1;\mp8i k \cos^{2}\frac{\sigma u}{2}\right),
&\left|\cos\frac{\sigma{u}}{2}\right|>1,
\vspace{3mm}\\
W_{1}(u)=e^{\pm4ik\cos^2\left(\sigma u/2\right)}\displaystyle \sum_{n=0}^{\infty}c_{n}^{(1)}
\hat{\Phi}\left(n+\frac{1}{2},n+1;\mp8i k \cos^{2}\frac{\sigma u}{2}\right),& \forall u
\end{array}
\end{eqnarray}
where the recurrence relations for the coefficients
are
\begin{eqnarray}
(n+1)c_{n+1}^{(1)}+
\left[n\left(n\pm8ik\right)\pm2ik+2k^2-a\right]c_{n}^{(1)}
\pm8ik \left(n-\frac{1}{2}\right)^2c_{n-1}^{(1)}=0
\end{eqnarray}
In these and in the following solutions the upper or
the lower sign must be taken throughout.

{\it Second set}. Even solutions, period $4\pi$ if $\sigma=1$
\antiletra\letra
\begin{eqnarray}\begin{array}{ll}
W_{2}^{0}(u)=\cos(\sigma u/2)\ e^{\pm4ik\cos^2(\sigma u/2)}
\displaystyle \sum_{n=0}^{\infty}
\frac{(-1)^{n}}{\Gamma[n+(1/2)]}c_{n}^{(2)}
\sin^{2n}\frac{\sigma u}{2},& \left|\cos\frac{\sigma{u}}{2}\right|<\infty,
\vspace{3mm}\\
W_{2}^{\infty}(u)=\cos(\sigma u/2)\ e^{\pm4ik\cos^2(\sigma u/2)}
\displaystyle \sum_{n=0}^{\infty}
c_{n}^{(2)}
\left(\mp8ik\cos^{2}\frac{\sigma u}{2}\right)^{-n-1},&
\left|\cos\frac{\sigma{u}}{2}\right|>1,
\vspace{3mm}\\
W_{2}(u)=\cos(\sigma u/2)\ e^{\pm4ik\cos^2(\sigma u/2)}
\displaystyle \sum_{n=0}^{\infty}c_{n}^{(2)}
\hat{\Phi}\left(n+1,n+2;\mp8ik\cos^{2}\frac{\sigma u}{2}\right),
& \forall u
\end{array}\end{eqnarray}
with the recurrence relations
\begin{eqnarray}
(n+1)c_{n}^{(2)}+
\left[n\left(n+1\pm8ik\right)+\frac{1}{4}\pm2ik+2k^2-a
\right]c_{n}^{(2)}
\pm 8ikn\left(n-\frac{1}{2}\right)c_{n-1}^{(2)}=0.
\end{eqnarray}
Note that in $W_{2}^{\infty}(u)$ we have eliminated the
function $\Psi\left(n+1,n+2;\mp8ik\cos^{2}(\sigma u/2)\right)$
by using the integral representation \cite{abramowitz}
\antiletra
\begin{eqnarray}\label{repint}
\Gamma(a)\Psi(a,b;x)= \int_{0}^{\infty}e^{-xt}t^{a-1}
(1+t)^{b-a-1}dt,\ \ (\Re{b}>\Re{a}>0)
\end{eqnarray}
along with the definite integral \cite{gradshteyn}
\begin{eqnarray}\label{defint}
 \int_{0}^{\infty}t^{n}e^{-\alpha t}dt=
\frac{\Gamma(n+1)}{\alpha^{n+1}}.
\end{eqnarray}
{\it Third set}. Odd solutions, period $2\pi$ if $\sigma=1$.
\letra
\begin{eqnarray}\begin{array}{ll}
W_{3}^{0}(u)=\sin(\sigma u)\ e^{\pm 4ik
\cos^2(\sigma u/2)}
\displaystyle\sum_{n=0}^{\infty}\frac{(-1)^{n}}
{\Gamma[n+(3/2)]}c_{n}^{(3)}
\sin^{2n}\frac{\sigma u}{2},&  \left|\cos\frac{\sigma{u}}{2}\right|<\infty,
\vspace{3mm}\\
W_{3}^{\infty}(u)=\sin(\sigma u)\ e^{\pm 4ik\cos^2(\sigma u/2)}
\displaystyle \sum_{n=0}^{\infty}
c_{n}^{(3)}
\Psi\left(n+\frac{3}{2},n+3;\mp 8ik
\cos^{2}\frac{\sigma u}{2}\right),& \left |\cos\frac{\sigma{u}}{2}\right|>1,
\vspace{3mm}\\
W_{3}(u)=\sin(\sigma u)\ e^{\pm 4ik\cos^2(\sigma u/2)}
\displaystyle \sum_{n=0}^{\infty}c_{n}^{(3)}
\hat{\Phi}\left(n+\frac{3}{2},n+3;\mp 8ik
\cos^{2}\frac{\sigma u}{2}\right),& \forall u
\end{array}\end{eqnarray}
having the recurrence relations
\begin{eqnarray}
 (n+1) c_{n+1}^{(3)}+
\left[n\left(n+2\pm 8ik\right)+1
\pm 6ik+2k^2-a\right]c_{n}^{(3)}
\pm 8ik\left(n+\frac{1}{2}\right)^{2}
c_{n-1}^{(3)}=0.
\end{eqnarray}

{\it Fourth set}. Odd solutions, period $4\pi$
if $\sigma=1$.
\antiletra\letra
\begin{eqnarray}\begin{array}{ll}
W_{4}^{0}(u)=\sin(\sigma u/2)\ e^{\pm 4ik \cos^2(\sigma u/2)}
\displaystyle \sum_{n=0}^{\infty}
\frac{(-1)^{n}}{\Gamma[n+(3/2)]}c_{n}^{(4)}
\sin^{2n}\frac{\sigma u}{2},& \left|\cos\frac{\sigma{u}}{2}\right|<\infty,
\vspace{3mm}\\
W_{4}^{\infty}(u)
=\sin(\sigma u/2)\ e^{\pm4i k\cos^2(\sigma u/2)}
\displaystyle \sum_{n=0}^{\infty}c_{n}^{(4)}
\left(\mp 8ik\cos^{2}\frac{\sigma u}{2}\right)^{-n-1},
& \left|\cos\frac{\sigma{u}}{2}\right|>1,
\vspace{3mm}\\
W_{4}(u)
=\sin(\sigma u/2)\ e^{\pm4 ik\cos^2(\sigma u/2)}
\displaystyle \sum_{n=0}^{\infty} c_{n}^{(4)}
\hat{\Phi}\left(n+\frac{3}{2},n+2;
\mp 8ik\cos^{2}\frac{\sigma u}{2}\right), & \forall u
\end{array}\end{eqnarray}
with
\begin{eqnarray}
 (n+1)c_{n+1}^{(4)}+
\left[ n\left(n+1\pm 8ik\right)+\frac{1}{4}
\pm 6ik+2k^2-a\right]c_{n}^{(4)}\pm 8ik n\left(n+\frac{1}{2}\right)c_{n-1}^{(4)}=0.
\end{eqnarray}

Finally, further solutions are obtained by
performing the substitution
\antiletra
\begin{eqnarray}
u\rightarrow u+\frac{\pi}{\sigma}
\end{eqnarray}
into the above solutions. This substitution, which leaves the
Mathieu equation (\ref{mathieu}) invariant, plays the role
of the transformation $T_{3}T_{4}$, for it transforms
$z=\cos^2(\sigma{u}/2)$ into $z=\sin^2(\sigma{u}/2)$
without changing $\omega$.
%
%
\section*{III. Solutions for the double-confluent Heun equation}
\indent
In the present section, first we obtain two sets of solutions for the DCHE
\begin{eqnarray*}
z^{2}\frac{d^{2}U}{dz^{2}}+
\left(B_{1}+B_{2}z\right)\frac{dU}{dz}+
\left(B_{3}-2\eta \omega z+\omega^{2}z^{2}\right)U=0,
\ \left(B_{1}\neq 0, \  \omega\neq 0\right),
\end{eqnarray*}
by taking the Leaver limit $z_{0}\rightarrow 0$ of
the first and third sets solutions for the GSWE given in
Sec II A. After examining the general case and discussing
finite-series solutions for a Schr\"odinger equations with
an asymmetric double-Morse potential,
we establish the properties of these
solutions for the Whittaker-Hill and Mathieu equations.

The following $U_{1}^{0}$ and $U_{2}^{0}$
are, in fact, the normal Thom\'e solutions in
the neighborhood of $z=0$. As mentioned before,
they encompass solutions which were investigated by Decareau,
Maroni and Robert \cite{decarreau2}
and also by Schmidt and Wolf \cite{schmidt}
for a DCHE with four parameters. However,
each of these solutions, resulting of the Baber-Hass\'e
expansions, are now associated with
two expansions in series of confluent hypergeometric
functions and these converge in the vicinity of $z=\infty$.
\subsection*{A. General case and solutions of Decarreau et al.}
\indent
When $z_{0}\rightarrow0$, the limits of the solutions
$U_{1}^{\infty}$ and $U_{1}$ given in Eq.
(\ref{frobenius1a}) are obtained immediately. The limit
of  $U_{1}^{0}$ yields the expression
\begin{eqnarray*}
U_{1}^{0}(z)=e^{i\omega z}\displaystyle \sum_{n=0}^{\infty}b_{n}^{(1)}z^{n},
\end{eqnarray*}
where the recurrence relations for $b_{n}^{(1)}$ are
\begin{eqnarray*}
B_{1}(n+1)b_{n+1}^{(1)}+
\left[ n\left(n+B_{2}-1\right)+
i\omega B_{1}+B_{3}\right]b_{n}^{(1)}
+2i\omega\left(n+i\eta+\frac{B_{2}}{2}-1\right)
b_{n-1}^{(1)}=0.
\end{eqnarray*}
Then, by writing $b_{n}^{(1)}=(B_{1})^{-n}c_{n}^{(1)}$,
we find the solutions given in Eqs. (\ref{soldche1}) and (\ref{recdche1}).
%

{\it First set}.
\letra
\begin{eqnarray}
\label{soldche1} \begin{array}{l}
U_{1}^{0}(z)=e^{i\omega z}\displaystyle \sum_{n=0}^{\infty}c_{n}^{(1)}
\left(\frac{z}{B_{1}}\right)^{n},
\vspace{3mm}\\
U_{1}^{\infty}(z)=e^{i\omega z}\displaystyle \sum_{n=0}^{\infty}c_{n}^{(1)}\Psi\left(n+i\eta+\frac{B_{2}}{2},n+B_{2};-2i\omega z\right),
\vspace{3mm}\\
U_{1}(z)=e^{i\omega z}
\displaystyle\sum_{n=0}^{\infty}c_{n}^{(1)}
\hat{\Phi}\left(n+i\eta+\frac{B_{2}}{2},n+B_{2},-2i\omega z\right),
\end{array}\end{eqnarray}
where the recurrence for relations $c_{n}^{(1)}$ are
($c_{-1}^{(1)}=0$)
\begin{eqnarray}\label{recdche1}
(n+1)c_{n+1}^{(1)}+
\left[ n\left(n+B_{2}-1\right)+
i\omega B_{1}+B_{3}\right]c_{n}^{(1)}
+2i\omega B_{1}\left(n+i\eta+\frac{B_{2}}{2}-1\right)
c_{n-1}^{(1)}=0.
\end{eqnarray}
%

%
%

{\it Second  set}. To obtain the Leaver limit of the solutions given in Eqs. (\ref{frobenius3}) we note that
\begin{eqnarray*}
z^{1+\frac{B_{1}}{z_{0}}}
(z-z_{0})^{1-B_{2}-\frac{B_{1}}{z_{0}}}=
z(z-z_{0})^{1-B_{2}}\left(1-\frac{z_{0}}{z}\right)^{-\frac{B_{1}}{z_{0}}}\rightarrow z^{2-B_{2}}e^{B_{1}/z},\ \ (z_{0}\rightarrow 0).
\end{eqnarray*}
Hence, replacing $b_{n}^{(3)}$
by $ (-B_{1})^{-n}c_{n}^{(3)}$,
solutions (\ref{frobenius3}) lead to
\antiletra\letra
\begin{eqnarray}\label{soldche2}\begin{array}{l}
U_{2}^{0}(z)=e^{i\omega z+(B_{1}/z)}z^{2-B_{2}}
\displaystyle \sum_{n=0}^{\infty}c_{n}^{(2)}
\left(-\frac{z}{B_{1}}\right)^{n},
\vspace{3mm}\\
U_{2}^{\infty}(z)=e^{i\omega z+(B_{1}/z)}z^{2-B_{2}}
\displaystyle \sum_{n=0}^{\infty}
c_{n}^{(2)}\Psi\left(n+2+i\eta-\frac{B_{2}}{2},
n+4-B_{2},-2i\omega z\right),
\vspace{3mm}\\
U_{2}(z)=e^{i\omega z+(B_{1}/z)}z^{2-B_{2}}
\displaystyle\sum_{n=0}^{\infty}c_{n}^{(2)}
\hat{\Phi}\left(n+2+i\eta-\frac{B_{2}}{2},n+4-B_{2},-2i\omega z\right),
\end{array}
\end{eqnarray}
with the recurrence relations
\begin{eqnarray}\label{recdche2}
&&(n+1)c_{n+1}^{(2)}
+\left[ n\left(n+3-B_{2}\right)+2-i\omega
 B_{1}-B_{2} +B_{3}\right]c_{n}^{(2)}
\nonumber\\
&&-2i\omega
B_{1}\left(n+1+i\eta-\frac{B_{2}}{2}
\right)c_{n-1}^{(2)}=0.
\end{eqnarray}

From the previous solutions we may obtain others
by the changes ($\eta,\omega$)$\leftrightarrow$
($-\eta,-\omega$).


{\it Schr\"odinger equation with asymmetric double-Morse potential}.
Now we consider the time-independent Schr\"{o}dinger equation
for particle with mass $m$ and energy $E$,
\antiletra
\begin{eqnarray}
\label{schr}
\frac{d^2\psi}{du^2}+[{\cal E}-V(u)]\psi=0, \ \ u:=\lambda x, \
 \ \ {\cal E}:=\frac{2mE}{\hbar^2 \lambda^2},
\end{eqnarray}
where $\lambda$ is a real constant,  $x$ is the spatial coordinate
and $V(u)$ is a function proportional to the potential. For the
asymmetric double-Morse potential considered by Zaslavskii and
Ulyanov \cite{zaslavskii,ulyanov} $V(u)$ is given by
\begin{eqnarray}
\label{zaslavskii2}
V(u)=\frac{B^2}{4}\left(\sinh{u}-\frac{C}{B}\right)^{2}
-B\left(s+\frac{1}{2}\right)\cosh{u}
\end{eqnarray}
where $B>0$, $C>0$ and $s$ is a non-negative
integer or half-integer. This is said to be a quasi-exactly solvable
potential because a part of the energy spectrum and
corresponding eigenfunctions can be found exactly
in closed form \cite{usheveridze1,usheveridze2}. Here
this means that one part of the energy spectrum is determined from
the recurrence relations of an eigenfunction
given by finite series. In section VI B, we show that
the other portion can be obtained from the recurrence
relations associated with an infinite-series eigenfunction.

For the potential (\ref{zaslavskii2}) the Schr\"odinger equation
assumes the form
\begin{eqnarray*}
\frac{d^2\psi}{du^2}+\left[{\cal E}-
\frac{C^2}{4}+\frac{B^2}{8}- \frac{B^2}{16}e^{-2u}-
\frac{B}{2}\left( \frac{C}{2}-\frac{1}{2}-s\right)e^{-u}+
\frac{B}{2}\left( \frac{C}{2}+\frac{1}{2}+s\right)e^{u}-
 \frac{B^2}{16}e^{2u}\right]\psi=0.
\end{eqnarray*}
The substitutions
\cite{eu2,lemieux}
\begin{eqnarray}\label{sol}
 z=e^{u},\
\psi(u)=\phi(z)=e^{-B/(4z)}z^{(C/2)-s} U(z),\ z\in[0,\infty)
\end{eqnarray}
transform the previous equation into the DCHE (\ref{dche})
\begin{eqnarray*}
z^{2}\frac{d^{2}U}{dz^{2}}+
\left[\frac{B}{2}+(1+C-2s)z\right]\frac{dU}{dz}+
\left[{\cal E}+\frac{B^2}{8}+s^2-Cs+\frac{B}{2}
\left( \frac{C}{2}+\frac{1}{2}+s\right)  z-
\frac{B^2}{16}z^{2}\right]U=0,
\end{eqnarray*}
and then we can choose the following set of parameters
\begin{eqnarray}
\label{paramzas}
B_{1}=\frac{B}{2}, \ B_{2}=1+C-2s,\
B_{3}={\cal E}+\frac{B^2}
{8}+s^2-sC, \
i\omega=-\frac{B}{4},\ i\eta=-\frac{C}{2}-\frac{1}{2}-s.
\end{eqnarray}
Hence, $\psi(u)=\phi(z)$ is constructed by inserting into
Eq.  (\ref{sol}) solutions of the DCHE  with
the parameters specified in Eq. (\ref{paramzas}) and
demanding that these eigenfuctions satisfy the regularity conditions
\begin{eqnarray}\label{regularity}
\lim_{z\to 0 }\phi(z)=0, \ \lim_{z\to\infty }\phi(z)=0.
\end{eqnarray}
Then, from the first set of solutions for the DCHE, we find
\begin{eqnarray}\label{sol1}
\begin{array}{l}
\phi_{1}^{0}(z) =z^{\frac{C}{2}-s}
\exp\left[-\frac{B}{4}\left(z+\frac{1}{z}\right)\right]
\displaystyle \sum_{n= 0}^{\infty}b_{n}^{(1)}
\left(\frac{2z}{B}\right)^{n},
\vspace{3mm}\\
\phi_{1}^{\infty}(z) =z^{\frac{C}{2}-s}
\exp\left[-\frac{B}{4}\left(z+\frac{1}{z}\right)\right]
\displaystyle \sum_{n= 0}^{\infty}b_{n}^{(1)}
\Psi\left(n-2s,n+1+C-2s;
\frac{Bz}{2}\right),
\end{array}
\end{eqnarray}
whose recurrence relations for $b_{n}^{(1)}$ are
\begin{eqnarray}\label{rec1}
(n+1)b_{n+1}^{(1)}+
\left[ n\left(n+C-2s\right)+{\cal E}+s^2-sC
\right]b_{n}^{(1)}
-\frac{B^2}{4}\left(n-2s-1\right)
b_{n-1}^{(1)}=0, \ \ b_{-1}^{(1)}=0.
\end{eqnarray}
Since $s$ is a non-negative
integer or half-integer, these are
finite-series solutions for the reason
given at the beginning of Section II.
In effect, the coefficient $\gamma_{n}^{(1)}$
of $b_{n-1}^{(1)}$ vanishes for $n=2s+1$ and
consequently the summation is restricted to the
range $0\leq n\leq 2s$. The regularity conditions are
satisfied by both solutions because $B>0$ in the exponential
factor; however only one solution is necessary to solve the
problem (the series of regular hypergeometric
functions is also appropriate). For a given value of $s$,
the energies may be found by computing the determinant
of the tridiagonal matrix (\ref{matriz}) corresponding
to Eq. (\ref{rec1}), with $0\leq n\leq 2s$.

First, notice that the above
solutions impose no additional restriction
on the parameter $C$, contrary to
the solutions employed in earlier approach\cite{eu2}.
Second, the solutions obtained from the other
set of solutions for the DCHE are given by infinite-series
but do not satisfy the regularity conditions: a suitable
two-sided solution ($-\infty<n<\infty$)
is presented in section VI. B.
Third,  for $C=0$ the potential becomes
a symmetric double-Morse potential
while the Schr\"odinger equation
reduces to the Whittaker-Hill equation. In
this case, by considering the WHE as a GSWE,
we find regular infinite-series solutions given
by one-sided series, that is, series with $n\geq 0$ \cite{eu2}.

%
\subsection*{B. Whittaker-Hill equation as a DCHE}
\indent
Here we find that the solutions for the Whittaker-Hill, considered
as a DCHE, possess no definite parity, contrary to the solutions which
result from the GSWE (Sec. II B).
Besides this, the periodicity (if any) will depend on the value of
the parameter $p$.

The substitutions
 \begin{eqnarray*}
z=e^{2i\kappa u}, \ \  W(u)=
z^{1+(p/2)}e^{\xi/(8z)}U(z)
\end{eqnarray*}
transform the WHE
\begin{eqnarray*}
\frac{d^2W}{du^2}+\kappa^2\left[\vartheta-\frac{1}{8}\xi^{2}
-(p+1)\xi\cos(2\kappa u)+
\frac{1}{8}\xi^{2}\cos(4\kappa u)\right]W=0,
\end{eqnarray*}
into
\begin{eqnarray*}
z^{2}\frac{d^{2}U}{dz^{2}}+
\left[-\frac{\xi}{4}+(p+3)z\right]\frac{dU}{dz}+
\left[\left(\frac{p}{2}+1\right)^{2}+\frac{\xi^2}{32}-
\frac{\vartheta}{4}+
\frac{(p+1)\xi}
{8} z-\frac{\xi^2}{64}z^{2}\right]U=0.
\end{eqnarray*}
Then, by comparing this with the DCHE (\ref{dche})
we conclude that the solutions $W(u)$ for the WHE are given by
\begin{eqnarray}\label{whe2}
\left.\begin{array}{l}
W(u)=z^{1+(p/2)}e^{\xi/(8z)}U(z), \ \ z=e^{2i\kappa u}\ (\kappa=1,i),
\vspace{3mm}\\
B_{1}=-\frac{\xi}{4}, \ B_{2}=p+3, \
B_{3}=\left(\frac{p}{2}+1\right)^{2}+\frac{\xi^2}{32}-
\frac{\vartheta}{4}, \ i\omega=\pm\frac{\xi}{8},
\ i\eta=\pm\frac{1}{2}(p+1)
\end{array}\right\}
\begin{array}{l}
\mbox{WHE  as}\vspace{3mm}\\
\mbox{DCHE,}
\end{array}
\end{eqnarray}
where now $U(z)$ denotes solutions for the DCHE
with the parameters specified in the second line of the above
expressions.

{\it First set}. Taking $i\omega=\xi/8$ and
$ i\eta=(p+1)/2$ and using the solutions
(\ref{soldche1}), we find
\letra
\begin{eqnarray}
\begin{array}{ll}
W_{1}^{0}(u)=e^{(\xi/4)
\cos(2\kappa u)+i(2+p)\kappa u}
\displaystyle \sum_{n=0}^{\infty}\left(-\frac{4}{\xi}\right)^{n}c_{n}^{(1)}
e^{2i \kappa n u}, &\left|e^ {2i\kappa u}\right|<\infty,
\vspace{3mm}\\
W_{1}^{\infty}(u)=e^{(\xi/4)
\cos(2\kappa u)+i(2+p)\kappa u}
\displaystyle \sum_{n=0}^{\infty}
\Psi\left(n+p+2,n+p+3;-\frac{\xi}{4}e^{2i\kappa u}\right),
&\left|e^ {2i\kappa u}\right|>0,
\vspace{3mm}\\
W_{1}(u)=e^{(\xi/4)
\cos(2\kappa u)+i(2+p)\kappa u}
\displaystyle \sum_{n=0}^{\infty}c_{n}^{(1)}
\hat{\Phi}
\left(n+p+2,n+p+3;-\frac{\xi}{4}e^{2i\kappa u}\right),
&\forall{u},
\end{array}
\end{eqnarray}
where the recurrence relations $c_{n}^{(1)}$ are
($c_{-1}^{(1)}=0$)
\begin{eqnarray}
(n+1)c_{n+1}^{(1)}+
\left[ n\left(n+p+2\right)+\left(\frac{p}{2}+1\right)^{2}
-\frac{\upsilon}{4}\right]c_{n}^{(1)}
-\frac{\xi^2}{16}\left(n+p+1\right)
c_{n-1}^{(1)}=0.
\end{eqnarray}
If $p$ is a negative integer equal or less than $-2$,
the series in these solutions are finite since the coefficient
of $c_{n-1}^{(1)}$ vanishes for some value of $n$.
For $\kappa=1$ the period of the solutions depends
on the factor $\exp{(i(p+2)u)}$. For instance,
the period is: $\pi$ if $p$ is an even integer;
$2\pi$ if $p$ is odd; $2m\pi$ ($m>1$) if $p=l/m$, where
$l$ and $m$ are prime to one another with $l<m$.

{\it Second set}. Taking again $i\omega=\xi/8$ and
$ i\eta=(p+1)/2$ but using the solutions
(\ref{soldche2}), we find
\antiletra\letra
\begin{eqnarray}
\begin{array}{ll}
W_{2}^{0}(u)=e^{(i\xi/4)
\sin(2\kappa u)-ip\kappa u}
\displaystyle \sum_{n=0}^{\infty}
\left(\frac{4}{\xi}
\right)^{n}c_{n}^{(2)}e^{2in \kappa u},
&\left|e^ {2i\kappa u}\right|<\infty,
\vspace{3mm}\\
W_{2}^{\infty}(u)=e^{(i\xi/4)
\sin(2\kappa u)-ip\kappa u}
\displaystyle\sum_{n=0}^{\infty}c_{n}^{(2)}
\Psi\left(n+1,n+1-p;-\frac{\xi}{4}
e^{2i\kappa u}\right),
&\left|e^ {2i\kappa u}\right|>0,
\vspace{3mm}\\
W_{2}(u)=e^{(i\xi/4)
\sin(2\kappa u)-ip\kappa u}
\displaystyle \sum_{n=0}^{\infty}c_{n}^{(2)}\hat{\Phi}
\left(n+1,n+1-p;-\frac{\xi}{4}
e^{2i\kappa u}\right), &\forall{u},
\end{array}
\end{eqnarray}
where the recurrence relations for $c_{n}^{(2)}$ are
\begin{eqnarray}
 (n+1)c_{n+1}^{(2)}+
\left[n\left(n-p\right)+
\frac{\xi^2}{16}+
\frac{p^2}{4}-\frac{\upsilon}{4}\right]
c_{n}^{(2)}+
\frac{\xi^2n}{16} c_{n-1}^{(2)}=0.
\end{eqnarray}
For these solutions the series are infinite.

{\it Third set}. Taking $i\omega=-\xi/8$ and
$ i\eta=-(p+1)/2$ and using the solutions
(\ref{soldche1}), we find
\antiletra\letra
\begin{eqnarray}
\begin{array}{ll}
W_{3}^{0}(u)=e^{-(i\xi/4)
\sin(2\kappa u)+i(p+2)\kappa u}
\displaystyle \sum_{n=0}^{\infty}
\left(-\frac{4}{\xi}
\right)^{n}c_{n}^{(3)}e^{2in \kappa u},
&\left|e^ {2i\kappa u}\right|<\infty,
\vspace{3mm}\\
W_{3}^{\infty}(u)=e^{-(i\xi/4)
\sin(2\kappa u)+i(p+2)\kappa u}
\displaystyle\sum_{n=0}^{\infty}c_{n}^{(3)}
\Psi\left(n+1,n+3+p;\frac{\xi}{4}
e^{2i\kappa u}\right),&\left|e^ {2i\kappa u}\right|>0,
\vspace{3mm}\\
W_{3}(u)=e^{-(i\xi/4)
\sin(2\kappa u)-ip\kappa u}
\displaystyle \sum_{n=0}^{\infty}c_{n}^{(3)}\hat{\Phi}
\left(n+1,n+3+p;\frac{\xi}{4}
e^{2i\kappa u}\right), &\forall{u},
\end{array}
\end{eqnarray}
where the recurrence relations for $c_{n}^{(3)}$ are
\begin{eqnarray}
 (n+1)c_{n+1}^{(3)}+
\left[n\left(n+2+p\right)+
\frac{\xi^2}{16}+
\left( 1+\frac{p}{2}\right) ^{2}-\frac{\upsilon}{4}\right]
c_{n}^{(3)}+
\frac{\xi^2n}{16} c_{n-1}^{(3)}=0.
\end{eqnarray}
This set can also be obtained from the second one by
$p\rightarrow -p-2$, $\xi\rightarrow -\xi$.
For these solutions the series are infinite.

{\it Fourth set}. Taking $i\omega=-\xi/8$ and
$ i\eta=-(p+1)/2$ and using the solutions
(\ref{soldche2}), we get
\antiletra\letra
\begin{eqnarray}\begin{array}{ll}
W_{4}^{0}(u)=e^{-(\xi/4)\cos(2\kappa u)-ip\kappa u}
\displaystyle \sum_{n=0}^{\infty}\left(\frac{4}{\xi}\right)^{n}c_{n}^{(4)}
e^{2i \kappa n u},&\left|e^ {2i\kappa u}\right|<\infty,
\vspace{3mm}\\
W_{4}^{\infty}(u)=e^{-(\xi/4)
\cos(2\kappa u)-ip\kappa u}
\displaystyle \sum_{n=0}^{\infty}c_{n}^{(4)}
\Psi\left(n-p,n+1-p;\frac{\xi}{4}e^{2i\kappa u}\right),
&\left|e^ {2i\kappa u}\right|>0,
\vspace{3mm}\\
W_{4}(u)=e^{-(\xi/4)
\cos(2\kappa u)-ip\kappa u}
\displaystyle \sum_{n=0}^{\infty}c_{n}^{(4)}
\hat{\Phi}
\left(n-p,n+1-p;\frac{\xi}{4}e^{2i\kappa u}\right),
&\forall{u},
\end{array}
\end{eqnarray}
where the recurrence relations $c_{n}^{(4)}$ are
\begin{eqnarray}
(n+1)c_{n+1}^{(4)}+
\left[ n\left(n-p\right)+\frac{p^2}{4}
-\frac{\upsilon}{4}\right]c_{n}^{(4)}
-\frac{\xi^2}{16}\left(n-p-1\right)
c_{n-1}^{(4)}=0.
\end{eqnarray}
This set also follows from the
first one by the changes $\xi\rightarrow -\xi$ and
$p\rightarrow -p-2$. If $p$ is a positive integer,
the series in these solutions are finite.

\subsection*{C. Mathieu equation as a DCHE}
\indent
The solutions $W(u)$ for the Mathieu equations (\ref{mathieu})
are obtained from the solutions given in Sec. III B for the WHE
by means of
\antiletra
\begin{eqnarray}
p=-1, \ 2\kappa=\sigma, \ \upsilon=4a-8k^2, \ \xi=\pm 8i k.
\end{eqnarray}

Unlike the solutions given in Sec. II.3, now there is no
solution with definite parity and for $\sigma=1$ ($u$ real)
the period is $4\pi$ instead of $2\pi$ and $4\pi$.
We get rid of the functions $\Psi$ which appear in $W_{1}^{\infty}$ and  $W_{2}^{\infty}$ by writing
\begin{eqnarray*}
\Psi\left(n+1,n+2; \pm2ike^{i\sigma{u}} \right)=
\left( \pm\frac{i}{2k}\right)^{n+1}e^{-i\sigma(n+1)u},
\end{eqnarray*}
where we have used equations (\ref{repint}) and (\ref{defint}).

{\it First set}. This comes from the first or fourth set of solutions
for the WHE.
\letra
\begin{eqnarray}
\begin{array}{ll}
W_{1}^{0}(u)=e^{\pm 2ik\cos(\sigma u)+
(i/2)\sigma u}
\displaystyle\sum_{n=0}^{\infty}
\left(\pm\frac{i}{2k}\right)^{n}c_{n}^{(1)}
e^{i \sigma n u},&\left|e^ {i\sigma u}\right|<\infty,
\vspace{3mm}\\
W_{1}^{\infty}(u)=e^{\pm 2ik\cos(\sigma u)-
(i/2)\sigma u}
\displaystyle \sum_{n=0}^{\infty}
\left(\pm\frac{i}{2k}\right)^{n}c_{n}^{(1)}
e^{-i\sigma n u},
&\left|e^ {i\sigma u}\right|>0,
\vspace{3mm}\\
W_{1}(u)=e^{\pm 2ik\cos(\sigma u)+
(i/2)\sigma u}
\displaystyle \sum_{n=0}^{\infty}c_{n}^{(1)}
\hat{\Phi}\left(n+1,n+2;\mp 2ik e^{i\sigma u}\right),
&\forall{u},
\end{array}
\end{eqnarray}
where the recurrence relations for $c_{n}^{(1)}$ are
($c_{-1}^{(1)}=0$)
\begin{eqnarray}
(n+1)c_{n+1}^{(1)}+
\left[ n\left(n+1\right)+\frac{1}{4}+2k^2
-a\right]c_{n}^{(1)}
+\left( 4k^2 n\right)  c_{n-1}^{(1)}=0.
\end{eqnarray}

{\it Second set}. This results from the second
or third set of solutions for the WHE.
\antiletra\letra
\begin{eqnarray}
\begin{array}{ll}
W_{2}^{0}(u)=e^{\mp 2k
\sin(\sigma u)+(i/2)\sigma u}
\displaystyle \sum_{n=0}^{\infty}
\left(\mp\frac{i}{2k}
\right)^{n}c_{n}^{(2)}e^{i \sigma nu},
&\left|e^ {i\sigma u}\right|<\infty,
\vspace{3mm}\\
W_{2}^{\infty}(u)=e^{\mp 2k
\sin(\sigma u)-(i/2)\sigma u}
\displaystyle\sum_{n=0}^{\infty}\left(\mp\frac{i}{2k}
\right)^{n}c_{n}^{(2)}
e^{-i\sigma {n}u},
&\left|e^ {i\sigma u}\right|>0,
\vspace{3mm}\\
W_{2}(u)=e^{\mp 2k
\sin(\sigma u)+(i/2)\sigma u}
\displaystyle \sum_{n=0}^{\infty}c_{n}^{(2)}\hat{\Phi}
\left(n+1,n+2;\mp 2ik
e^{i\sigma u}\right),&\forall{u},
\end{array}
\end{eqnarray}
where the recurrence relations for $c_{n}^{(2)}$ are ($c_{-1}^{(2)}=0$)
\begin{eqnarray}
 (n+1)c_{n+1}^{(2)}+
\left[n\left(n+1\right)
+\frac{1}{4}-2k^2-a\right]
c_{n}^{(2)}-\left( 4k^2 n\right) c_{n-1}^{(2)}=0.
\end{eqnarray}
%
\section*{IV. Whittaker-Ince's limits for the GSWE}
\indent
Four sets of solutions for the Whittaker-Ince limit (\ref{incegswe})
of the GSWE, that is, for
\antiletra
\begin{eqnarray}\label{incegswe1}
z(z-z_{0})\frac{d^{2}U}{dz^{2}}+(B_{1}+B_{2}z)
\frac{dU}{dz}+
\left[B_{3}+q(z-z_{0})\right]U=0,\ (q\neq0),
\end{eqnarray}
result from the solutions given in Sec. II A by means
of the very Whittaker-Ince limit ($\omega\rightarrow0$,
$\eta\rightarrow\infty$ such that $2\eta\omega=-2q$).
However, we consider only the limit of the first set and, then, use
transformation rules to obtain the other sets of solutions.
In effect, if $U(z)=U(B_{1},B_{2},B_{3};
z_{0},q;z)$ denotes one solution for Eq. (\ref{incegswe1}),
the others are obtained by using the rules
\begin{eqnarray}\label{trans} \begin{array}{l}
T_{1}U(z)=z^{1+B_{1}/z_{0}}
U(C_{1},C_{2},C_{3};z_{0},q;z),\  \ z_{0}\neq0,
\vspace{3mm}\\
T_{2}U(z)=(z-z_{0})^{1-B_{2}-B_{1}/z_{0}}U(B_{1},D_{2},D_{3};
z_{0},q;z), \ \  z_{0}\neq0,
\vspace{3mm}\\
T_{4}U(z)=
U(-B_{1}-B_{2}z_{0},B_{2},
B_{3}-q z_{0};z_{0},-q;z_{0}-z),
\end{array}
\end{eqnarray}
where $C_{i}$ and $D_{i}$ are defined as section II.
We again use only $T_{1}$ and
$T_{2}$ but notice  that $T_{4}$ may be used to get
alternative representations for the solutions. As a
particular case we set up solutions for the Mathieu equation.
\subsection*{A. General case}
\indent
First we rewrite the solutions given in Eq. (\ref{frobenius1a}) as
\begin{eqnarray*} \begin{array}{l}
U_{1}^{0}(z)=e^{i\omega z}\displaystyle \sum_{n=0}^{\infty}b_{n}^{(1)}
(z-z_{0})^n,
\vspace{3mm}\\
U_{1}^{\infty}(z)=e^{i\omega z}\displaystyle \sum_{n=0}^{\infty}c_{n}^{(1)}
\Gamma\left[1+i\eta-\frac{B_{2}}{2}\right]
\Psi\left(n+i\eta+\frac{B_{2}}{2},n+B_{2};-\frac{qz}{i\eta}\right),
\vspace{3mm}\\
U_{1}(z)=e^{i\omega z}\displaystyle
\sum_{n=0}^{\infty} c_{n}^{(1)}
\hat{\Phi}\left(n+i\eta+\frac{B_{2}}{2},n+B_{2};
-\frac{qz}{i\eta}\right),
\end{array}\end{eqnarray*}
where we have multiplied the solution $U_{1}^{\infty}(z)$
by $\Gamma\left[1+i\eta-\frac{B_{2}}{2}\right]$ and written
$q=-2\eta \omega$.
From these solutions and respective
recurrence relations for their coefficients,
the limits are found by considering $n$ fixed ($a\sim i\eta$)
and using the formulae \cite{erdelyi1}
\begin{eqnarray}\label{K}\begin{array}{l}
\displaystyle
\lim_{a\rightarrow  \infty}\Phi\left(a,b;-\frac{x}{a}\right)=
\Gamma(b)x^{(1-b)/2}J_{b-1}(2\sqrt{x}),\vspace{3mm}\\
\displaystyle
\lim_{a\rightarrow  \infty}\left[\Gamma(a+1-b)\Psi\left(a,b;\frac{x}{a}\right)\right]=
2x^{(1-b)/2}K_{b-1}(2\sqrt{x})
\end{array}
\end{eqnarray}
where $J_{\lambda}(\xi)$ and $K_{\lambda}(\xi)$ denote the
Bessel and the modified Bessel function of the second
kind \cite{luke}, whose definitions in terms of
confluent hypergeometric functions
are \cite{erdelyi1}
\begin{eqnarray}\label{2.8}\begin{array}{l}
J_{\lambda}(t)=\frac{1}{\Gamma(\lambda+1)}
e^{-it}\left(\frac{t}{2}\right)^{\lambda}\Phi\left(\lambda+\frac{1}{2},
2\lambda+1;2it\right),\vspace{3mm}\\
K_{\lambda}(t)=K_{-\lambda}(t)=\sqrt{\pi}\ e^{-t}(2t)^{\lambda}
\Psi\left(\lambda+\frac{1}{2},2\lambda+1,2t\right).
\end{array}
\end{eqnarray}
For using (\ref{K}) when $i\eta\rightarrow \infty$
($n$ and $B_{2}$ fixed, $q=$constant), first we
make the approximations
\begin{eqnarray*}
&&\Psi\left(n+i\eta+\frac{B_{2}}{2},n+B_{2};-\frac{qz}{i\eta}\right)\approx \Psi\left(i\eta,n+B_{2};-\frac{qz}{i\eta}\right),\\
&&\hat{\Phi}\left(n+i\eta+\frac{B_{2}}{2},n+B_{2};-\frac{qz}{i\eta}\right)\approx \hat{\Phi}\left(i\eta,n+B_{2};-\frac{qz}{i\eta}\right).
\end{eqnarray*}
In this manner we find the first set of solutions written
in Eq. (\ref{limitleaver1}). Notice that, although this is a formal derivation,
the solutions can be tested by inserting them into
Eq. (\ref{incegswe1}), as shown in Appendix A.

%
{\it First set}. These solutions admit the Leaver limit. If $B_{2}+B_{1}/z_{0}=1$, the first and the fourth sets of solutions are equal to one another.
\letra
\begin{eqnarray}\label{limitleaver1}
\begin{array}{l}
U_{1}^{0}(z)=\displaystyle \sum_{n=0}^{\infty}b_{n}^{(1)}
(z-z_{0})^{n},
\vspace{3mm}\\
U_{1}^{\infty}(z)=z^{(1-B_{2})/2}
\displaystyle \sum_{n=0}^{\infty}
c_{n}^{(1)}(i\sqrt{qz})^{-n}K_{n+B_{2}-1}(2i\sqrt{qz}),
\vspace{3mm}\\
U_{1}(z)=z^{(1-B_{2})/2}
\displaystyle \sum_{n=0}^{\infty}
c_{n}^{(1)}(\sqrt{qz})^{-n}J_{n+B_{2}-1}(2\sqrt{qz}),
\end{array}\end{eqnarray}
where the recurrence relations for $b_{n}^{(1)}$ and $c_{n}^{(1)}$ are
($b_{-1}^{(1)}=c_{-1}^{(1)}=0$)
\begin{eqnarray}
z_{0}\left(n+B_{2}+\frac{B_{1}}{z_{0}}\right)(n+1)b_{n+1}^{(1)}+
\left[ n\left(n+B_{2}-1\right)+B_{3}\right]
b_{n}^{(1)}+qb_{n-1}^{(1)}=0,
\end{eqnarray}
and
\begin{eqnarray}
(n+1)c_{n+1}^{(1)}+
\left[ n\left(n+B_{2}-1\right)+B_{3}\right]
c_{n}^{(1)}+qz_{0}\left(n+B_{2}+\frac{B_{1}}{z_{0}}-1\right)c_{n-1}^{(1)}=0.
\end{eqnarray}
The coefficients $b_{n}^{(1)}$ and $c_{n}^{(1)}$
are again connected by Eq. (\ref{connect1}). From
this first set of solutions we obtain the others by applying
the previous rules $T_{1}$ and $T_{2}$
according to relations (\ref{rules}).
%

%
%
%
%
%
%
{\it Second set}. These solutions do not admit the Leaver limit.
If $B_{2}+B_{1}/z_{0}=1$, the second and the third sets are
equal to each other.
\antiletra
\letra
\begin{eqnarray}\begin{array}{l}
U_{2}^{0}(z)=z^{1+(B_{1}/z_{0})}\displaystyle \sum_{n=0}^{\infty}b_{n}^{(2)}(z-z_{0})^{n},
\vspace{3mm}\\
U_{2}^{\infty}(z)=z^{(1-B_{2})/2}\displaystyle \sum_{n=0}^{\infty}c_{n}^{(2)}
(i\sqrt{qz})^{-n} K_{n+1+B_{2}+\frac{2B_{1}}{z_{0}}}( 2i\sqrt{qz}),
\vspace{3mm}\\
U_{2}(z)=z^{(1-B_{2})/2}\displaystyle \sum_{n=0}^{\infty}c_{n}^{(2)}
(\sqrt{qz})^{-n} J_{n+1+B_{2}+\frac{2B_{1}}{z_{0}}}( 2\sqrt{qz}),
\end{array}\end{eqnarray}
where the recurrence relations are given by
\begin{eqnarray}
&&z_{0}\left(n+B_{2}+\frac{B_{1}}{z_{0}}\right)(n+1)b_{n+1}^{(2)}+
\vspace{3mm}\nonumber\\
&&\left[n\left(n+1+B_{2}+\frac{2B_{1}}{z_{0}}\right)+
\left(1+\frac{B_{1}}{z_{0}}\right)\left(B_{2}+
\frac{B_{1}}{z_{0}}\right)+B_{3}\right]b_{n}^{(2)}
+qb_{n-1}^{(2)}=0
\end{eqnarray}
and
\begin{eqnarray}
&&(n+1)c_{n+1}^{(2)}+
\left[n\left(n+1+B_{2}+\frac{2B_{1}}{z_{0}}\right)+
\left(1+\frac{B_{1}}{z_{0}}\right)\left(B_{2}+
\frac{B_{1}}{z_{0}}\right)+B_{3}\right]c_{n}^{(2)}\vspace{3mm}+\nonumber\\
&&qz_{0}\left(n+B_{2}+\frac{B_{1}}{z_{0}}-1\right)c_{n-1}^{(2)}=0.
\end{eqnarray}
The relation between $b_{n}^{(2)}$ and $c_{n}^{(2)}$
is given by Eq. (\ref{connect2}).
%

%
%
{\it Third set}. These solutions admit Leaver limit.
\antiletra
\letra
\begin{eqnarray}\label{limitleaver2}
\begin{array}{l}
U_{3}^{0}(z)=z^{1+\frac{B_{1}}{z_{0}}}
(z-z_{0})^{1-B_{2}-\frac{B_{1}}{z_{0}}}
\displaystyle \sum_{n=0}^{\infty}b_{n}^{(3)}(z-z_{0})^{n},
\vspace{3mm}\\
U_{3}^{\infty}(z)=z^{\frac{B_{2}-1}{2}+\frac{B_{1}}{z_{0}}}
(z-z_{0})^{1-B_{2}-\frac{B_{1}}{z_{0}}}
\displaystyle \sum_{n=0}^{\infty}c_{n}^{(3)}(i \sqrt{qz})^{-n}
K_{n+3-B_{2}}(2i\sqrt{qz}),
\vspace{3mm}\\
U_{3}(z)=z^{\frac{B_{2}-1}{2}+\frac{B_{1}}{z_{0}}}
(z-z_{0})^{1-B_{2}-\frac{B_{1}}{z_{0}}}
\displaystyle \sum_{n=0}^{\infty}c_{n}^{(3)}
( \sqrt{qz})^{-n}
J_{n+3-B_{2}}(2\sqrt{qz}),
\end{array}
\end{eqnarray}
with the recurrence relations
\begin{eqnarray}
z_{0}\left(n+2-B_{2}-\frac{B_{1}}{z_{0}}\right)
(n+1)b_{n+1}^{(3)}+\left[n\left(n+3
-B_{2}\right)+2-B_{2}+B_{3}
\right]b_{n}^{(3)}+qb_{n-1}^{(3)}=0
\end{eqnarray}
and
\begin{eqnarray}
(n+1)c_{n+1}^{(3)}+\left[n\left(n+3
-B_{2}\right)+2-B_{2}+B_{3}
\right]c_{n}^{(3)}+qz_{0}\left(n+1-B_{2}-\frac{B_{1}}{z_{0}}
\right)c_{n-1}^{(3)}=0.
\end{eqnarray}
Eq. (\ref{connect3}) affords the relations between $b_{n}^{(3)}$ and $c_{n}^{(3)}$.

{\it Fourth set}. There is no limit for $z_{0}\rightarrow 0$.
\antiletra
\letra
\begin{eqnarray}\begin{array}{l}
U_{4}^{0}(z)=(z-z_{0})^{1-B_{2}-\frac{B_{1}}{z_{0}}}
\displaystyle \sum_{n=0}^{\infty}b_{n}^{(4)}(z-z_{0})^{n},
\vspace{3mm}\\
U_{4}^{\infty}(z)=z^{\frac{B_{1}}{z_{0}}+
\frac{B_{2}}{2}-\frac{1}{2}}(z-z_{0})^{1-B_{2}-\frac{B_{1}}{z_{0}}}
\displaystyle \sum_{n=0}^{\infty}c_{n}^{(4)}(i\sqrt{qz})^{-n}
K_{n+1-B_{2}-\frac{2B_{1}}{z_{0}}}(2i\sqrt{qz}),
\vspace{3mm}\\
U_{4}(z)=z^{\frac{B_{1}}{z_{0}}+
\frac{B_{2}}{2}-\frac{1}{2}}(z-z_{0})^{1-B_{2}-\frac{B_{1}}{z_{0}}}
\displaystyle \sum_{n=0}^{\infty}c_{n}^{(4)}(\sqrt{qz})^{-n}
J_{n+1-B_{2}-\frac{2B_{1}}{z_{0}}}(2\sqrt{qz})
\end{array}
\end{eqnarray}
with the recurrence relations
\begin{eqnarray}
&&z_{0}\left(n+2-B_{2}-\frac{B_{1}}{z_{0}}\right)(n+1)b_{n+1}^{(4)}+
\nonumber\\
&&\left[n\left(n+1-B_{2}-
\frac{2B_{1}}{z_{0}}\right)+
\frac{B_{1}}{z_{0}}\left(B_{2}+\frac{B_{1}}{z_{0}}-1\right)
+B_{3}\right]
b_{n}^{(4)}+qb_{n-1}^{(4)}=0
\end{eqnarray}
and
\begin{eqnarray}
&&(n+1)c_{n+1}^{(4)}+
\left[n\left(n+1-B_{2}-
\frac{2B_{1}}{z_{0}}\right)+
\frac{B_{1}}{z_{0}}\left(B_{2}+\frac{B_{1}}{z_{0}}-1\right)
+B_{3}\right]
c_{n}^{(4)}+\nonumber\vspace{3mm}\\
&&qz_{0}\left(n+1-B_{2}-\frac{B_{1}}{z_{0}}\right)
c_{n-1}^{(4)}=0.
\end{eqnarray}
Now the coefficients $b_{n}^{(4)}$ and $c_{n}^{(4)}$
are connected by Eq. (\ref{connect4}).

For the aforementioned wave equation
which arises from the separation of
variables of the Laplace-Beltrami operator for a scalar field
on the Eguchi-Hanson space, Malmendier \cite{malmendier}
has also found a solution in power series
whose coefficients satisfy three-term recurrence relations.
However, as a solution convergent in the neighborhood
of $z=\infty$, he has proposed a subnormal Thom\'e solution
with four-term recurrence relations, in contrast with
the above solutions in series of Bessel functions.
\subsection*{B. Mathieu equation and Lindemann-Stieltjes' solutions }
\indent
By inserting
\begin{eqnarray*}
z_{0}=1,\ B_{1}=-1/2, \ B_{2}=1,\ z=\cos^2{(\sigma u)},\
W(u)=U(z),
\end{eqnarray*}
into Eq. (\ref{incegswe1}) we obtain the Mathieu equation
\begin{eqnarray*}
\frac{d^2W}{du^2}+\sigma^2\left[2q-4B_{3}-2q
\cos(2\sigma u)\right]W=0.
\end{eqnarray*}
Then, solutions for the Mathieu equation are obtained by making
\antiletra
\begin{eqnarray}\label{mathieu4}
\left.\begin{array}{l}
W(u)=U(z), \ \ z=\cos^{2}(\sigma u),\
( \sigma=1,i),
\vspace{3mm}\\
z_{0}=1,\ B_{1}=-{1}/{2}, \ B_{2}=1, \
B_{3}=\frac{1}{4}[2q-a],\ q=k^2,
\end{array}\right\}
\begin{array}{l}
\mbox{Mathieu Eq. as Whittaker- }\vspace{3mm}\\
\mbox{Ince limit of the GSWE},
\end{array}
\end{eqnarray}
where $\sigma=1$ for the Mathieu equation, $\sigma=i$
for the modified Mathieu equation and $U(z)$ are the
solutions for the Whittaker-Ince limit of the GSWE, Sec. IV A.
Amongst the following solutions, included in the $W_{i}^{0}(u)$,
are solutions found by Lindmann and Stieltjes, which
are even or odd and have period $\pi$ or $2\pi$ unlike
the solutions of Sections II C and III C. As mentioned at
the beginning of Sec II A, the following solutions may also
be generated by letting $\xi$ and $p$ to tend respectively
to zero and infinity so that $p\xi=2k^{2}$
in the  solutions of Sec. II B to WHE (\ref{whe}) ($\kappa=\sigma$),
that is, by following  the original proposal of
Whittaker and Ince \cite{humbert,ince,ince2}.

{\it First set}. Even solutions. The solutions
$W_{1}^{0}$ and $W_{1}$ have period $\pi$ if
$\sigma=1$ since $J_{n}(-x)=(-1)^nJ_{n}(x)$.
\letra
\begin{eqnarray}\begin{array}{ll}
W_{1}^{0}(u)=\displaystyle \sum_{n=0}^{\infty}
\frac{(-1)^{n}c_{n}^{(1)}}{\Gamma[n+(1/2)]}
\sin^{2n}(\sigma u),&|\cos(\sigma u)|<\infty,
\vspace{3mm}\\
W_{1}^{\infty}(u)=
\displaystyle \sum_{n=0}^{\infty}
c_{n}^{(1)}\left[ik\cos(\sigma u)\right]^{-n}
K_{n}\left[2ik\cos(\sigma u)\right],&|\cos(\sigma u)|>1,
\vspace{3mm}\\
W_{1}(u)=
\displaystyle \sum_{n=0}^{\infty}
c_{n}^{(1)}\left[k\cos(\sigma u)\right]^{-n}
J_{n}\left[2k\cos(\sigma u)\right],&\forall u.
\end{array}\end{eqnarray}
The recurrence relations are
\begin{eqnarray}
(n+1)c_{n+1}^{(1)}+
\left( n^2+\frac{k^2}{2}-\frac{a}{4}\right)
c_{n}^{(1)}+k^2 \left(n-\frac{1}{2}\right)c_{n-1}^{(1)}=0.
\end{eqnarray}
{\it Second set}. Even solutions. If $\sigma=1$  $W_{2}^{0}$ and $W_{2}$  have period $2\pi$.
\antiletra\letra
\begin{eqnarray}\begin{array}{ll}
W_{2}^{0}(u)=\cos(\sigma u)\displaystyle \sum_{n=0}^{\infty}\frac{(-1)^nb_{n}^{(1)}}{\Gamma[n+(1/2)]}
\sin^{2n}(\sigma u),&|\cos(\sigma u)|<\infty,
\vspace{3mm}\\
W_{2}^{\infty}(u)=
\displaystyle \sum_{n=0}^{\infty}
c_{n}^{(1)}\left[ik\cos(\sigma u)\right]^{-n}
K_{n+1}\left[2ik\cos(\sigma u)\right],&|\cos(\sigma u)|>1,
\vspace{3mm}\\
W_{2}(u)=
\displaystyle \sum_{n=0}^{\infty}
c_{n}^{(1)}\left[k\cos(\sigma u)\right]^{-n}
J_{n+1}\left[2k\cos(\sigma u)\right],&\forall u,
\end{array}\end{eqnarray}
with the recurrence relations
\begin{eqnarray}
(n+1)c_{n+1}^{(2)}+
\left[ n(n+1)+\frac{1}{4}+\frac{k^2}{2}
-\frac{a}{4}\right]
c_{n}^{(2)}+k^2 \left(n-\frac{1}{2}\right)c_{n-1}^{(2)}=0.
\end{eqnarray}
%
{\it Third set}. Odd solutions. If $\sigma=1$, $W_{3}^{0}$ and $W_{3}$ have period $\pi$.
\antiletra\letra
\begin{eqnarray}\begin{array}{ll}
W_{3}^{0}(u)=\cos(\sigma u)\displaystyle \sum_{n=0}^{\infty}
\frac{(-1)^nc_{n}^{(3)}}{\Gamma[n+(1/2)]}
\sin^{2n+1}(\sigma u),&|\cos(\sigma u)|<\infty,
\vspace{3mm}\\
W_{3}^{\infty}(u)=
\tan(\sigma u)\displaystyle \sum_{n=0}^{\infty}
c_{n}^{(3)}\left[ik\cos(\sigma u)\right]^{-n}
K_{n+2}\left[2ik\cos(\sigma u)\right],&|\cos(\sigma u)|>1,
\vspace{3mm}\\
W_{3}(u)=
\tan(\sigma u)\displaystyle \sum_{n=0}^{\infty}
c_{n}^{(3)}\left[k\cos(\sigma u)\right]^{-n}
J_{n+2}\left[2k\cos(\sigma u)\right],&|\forall u.
\end{array}\end{eqnarray}
The recurrence relations are
\begin{eqnarray}
(n+1)c_{n+1}^{(3)}+
\left[ n(n+2)+1+\frac{k^2}{2}-\frac{a}{4}\right]
c_{n}^{(3)}+k^2 \left(n+\frac{1}{2}\right)c_{n-1}^{(3)}=0.
\end{eqnarray}

%
%
{\it Fourth set}. Odd solutions. For
$\sigma=1$, $W_{4}^{0}(z)$
and $W_{4}(z)$ have period $2\pi$.
\antiletra\letra
\begin{eqnarray}\begin{array}{ll}
W_{4}^{0}(u)=\displaystyle \sum_{n=0}^{\infty}
\frac{(-1)^nb_{n}^{(4)}}{\Gamma[n+(3/2)]}
\sin^{2n+1}(\sigma u),&|\cos(\sigma u)|<\infty,
\vspace{3mm}\\
W_{4}^{\infty}(u)=\tan(\sigma u)
\displaystyle \sum_{n=0}^{\infty}
c_{n}^{(4)}\left[ik\cos(\sigma u)\right]^{-n}
K_{n+1}\left[2ik\cos(\sigma u)\right],&|\cos(\sigma u)|>1,
\vspace{3mm}\\
W_{4}(u)=\tan(\sigma u)
\displaystyle \sum_{n=0}^{\infty}
c_{n}^{(4)}\left[k\cos(\sigma u)\right]^{-n}
J_{n+1}\left[2k\cos(\sigma u)\right],&\forall u.
\end{array}\end{eqnarray}
The recurrence relations are
\begin{eqnarray}
(n+1)c_{n+1}^{(4)}+
\left[ n(n+1)+\frac{1}{4}+
\frac{k^2}{2}-\frac{a}{4}\right]c_{n}^{(4)}+k^2 \left(n+\frac{1}{2}\right)c_{n-1}^{(4)}=0.
\end{eqnarray}

From the preceding solutions other expansions may be
obtained by the substitutions
\antiletra
\begin{eqnarray}\label{stieltjes}
u\rightarrow u+\frac{\pi}{2\sigma}\ \mbox{and}\ k\rightarrow ik,
\end{eqnarray}
which plays the role of the transformation
$T_{4}$ given in Eq. (\ref{trans}). The solutions
$W_{1}^{0}$, $W_{2}^{0}$ and those resulting
from these by means of (\ref{stieltjes}), if expressed
in terms of $z$ [$z=\cos^2{(\sigma{u})}$], are the ones
proposed by Lindemann and Stieltjes
\cite{lindemann,watson} but here to each of these
now correspond two solutions in series of Bessel
functions.
%
%
\section*{V. Whittaker-Ince's limits for the DCHE}
\indent
The solutions for the Whittaker-Ince limit  (\ref{incedche})
of the DCHE, that is, for
\begin{eqnarray}\label{incedche2}
z^2\frac{d^{2}U}{dz^{2}}+(B_{1}+B_{2}z)
\frac{dU}{dz}+
\left(B_{3}+qz\right)U=0,\ (q\neq0, \ B_{1}\neq 0)
\end{eqnarray}
may be obtained from the solutions for the DCHE,
given Sec. III A, by letting $\omega\rightarrow0$
and $\eta\rightarrow\infty$ such that $2\eta\omega=-q$.
However, at this stage, it is simpler to take the Leaver
limit $z_{0}\rightarrow0$ of the first and third sets of
solutions given in Sec. IV A. As in the cases of the
Mathieu equation there is no finite-series solution.
%

{\it First set}. If we put $b_{n}^{(1)}=c_{n}^{(1)}/(B_{1})^{n}$
in the limits of solutions (\ref{limitleaver1}), we find
\letra
\begin{eqnarray}\begin{array}{l}
U_{1}^{0}(z)=\displaystyle \sum_{n=0}^{\infty}c_{n}^{(1)}
\left(\frac{z}{B_{1}}\right)^{n},
\vspace{3mm}\\
U_{1}^{\infty}(z)=z^{(1-B_{2})/2}
\displaystyle \sum_{n=0}^{\infty}
c_{n}^{(1)}(i\sqrt{qz})^{-n}K_{n+B_{2}-1}(2i\sqrt{qz}),
\vspace{3mm}\\
U_{1}(z)=z^{(1-B_{2})/2}
\displaystyle \sum_{n=0}^{\infty}
c_{n}^{(1)}(\sqrt{qz})^{-n}J_{n+B_{2}-1}(2\sqrt{qz}),
\end{array}\end{eqnarray}
where the recurrence relations for $c_{n}^{(1)}$ are
($c_{-1}^{(1)}=0$)
\begin{eqnarray}
(n+1)c_{n+1}^{(1)}+
\left[ n\left(n+B_{2}-1\right)+B_{3}\right]
c_{n}^{(1)}+qB_{1}c_{n-1}^{(1)}=0.
\end{eqnarray}
%
%
%
{\it Second set}. In the limit of third set of solutions given in Sec. IV A
we take
\begin{eqnarray*}
b_{n}^{(3)}=c_{n}^{(2)}/(-B_{1})^{n}, \ \
\displaystyle\lim_{z_{0}\rightarrow 0}
\left(1-\frac{z_{0}}{z}\right)^{-B_{1}/z_{0}}=
e^{B_{1}/z}.
\end{eqnarray*}
Then, we obtain
\antiletra
\letra
\begin{eqnarray}\begin{array}{l}
U_{2}^{0}(z)=z^{2-B_{2}}
e^{B_{1}/z}
\displaystyle \sum_{n=0}^{\infty}c_{n}^{(2)}\left(-\frac{z}{B_{1}}\right)^{n},
\vspace{3mm}\\
U_{2}^{\infty}(z)=z^{(1-B_{2})/2}
e^{B_{1}/z}
\displaystyle \sum_{n=0}^{\infty}c_{n}^{(2)}(i \sqrt{qz})^{-n}
K_{n+3-B_{2}}(2i\sqrt{qz}),
\vspace{3mm}\\
U_{2}(z)=z^{(1-B_{2})/2}e^{B_{1}/z}
\displaystyle \sum_{n=0}^{\infty}c_{n}^{(2)}
( \sqrt{qz})^{-n}
J_{n+3-B_{2}}(2\sqrt{qz}),
\end{array}
\end{eqnarray}
with the recurrence relations
\begin{eqnarray}
(n+1)c_{n+1}^{(2)}+\left[n\left(n+3
-B_{2}\right)+2-B_{2}+B_{3}
\right]c_{n}^{(2)}-qB_{1}c_{n-1}^{(2)}=0.
\end{eqnarray}
As we see, in the present context, it is almost trivial to
obtain solutions for Eq. (\ref{incedche2}).
\section*{VI. Solutions with a phase parameter}

In this section we make some comments on solutions in
which a phase parameter $\nu$ was introduced to assure
the series convergence if there is no free
constant in the GSWE. Such solutions were
partly studied elsewhere \cite{leaver1,eu2,otchik1,mano1} but
here we provide some more informations useful in
applications. These solutions, which have been used
in the study of the Teukolsky equations in Kerr spacetimes
\cite{otchik1,mano1,mano2,mano3}, will as well be
taken as a possible starting-point to find
new solutions to the general Heun equation (Appendix B).

Instead of three solutions as in Ref. \cite{eu2}, we consider
a basic set containing four solutions, two given by expansions in
series of hypergeometric functions $F(a,b;c;y)$
and two by expansions in series of confluent
hypergeometric functions. These solutions, like the ones of
Sec. II, admit both the Leaver and
the Whittaker-Ince limits, but the last limit is not regarded here
because  it is the same as in  Ref. \cite{eu}, apart from the
fact that now we have four solutions instead of two.

In the first place we remark that there are several physical
problems ruled by differential equations without any free
constant, that is, requiring a phase parameter. For example,
all the constants
appearing in the GSWEs responsible by the time-dependence
of Klein-Gordon and Dirac test-fields in some curved
spacetimes are determined from the spatial part of
the wavefunction \cite{eu2,scho1,scho2,birrel}. Similarly,
in the radial Schr\"odinger equation for  the
scattering of electrons by a finite dipole \cite{leaver1}
or by polarizable targets \cite{eu,kleinman,buhring2,buhring1}
there is no disposable constant, since
all the parameters which characterize the target and
the incident particles are known.

In the second place, solutions
with a phase parameter $\nu$
are given by doubly infinite series ($-\infty <n <\infty$) and,
if $b_{n}$ denotes the series coefficients, then the
recurrence relations have the general form
\antiletra\letra
\begin{eqnarray}\label{twosided}
\alpha_{n}b_{n+1}+\beta_{n}b_{n}+\gamma_{n}b_{n-1}=0,\ (-\infty<n<\infty)
\end{eqnarray}
where $\alpha_{n}$, $\beta_{n}$, $\gamma_{n}$
and $b_{n}$ depend on $\nu$ and on the parameters of the
GSWE. These recurrence relations give the characteristic
equation \cite{leaver1}
\begin{eqnarray}\label{twosidedcharac}
\beta_{0}=\frac{\alpha_{-1}\gamma_{0}}{\beta_{-1}-} \frac{\alpha_{-2}
\gamma_{-1}}{\beta_{-2}-}\frac{\alpha_{-3}\gamma_{-2}}
{\beta_{-3}-}\cdots+\frac{\alpha_{0}\gamma_{1}}{\beta_{1}-}
\frac{\alpha_{1}\gamma_{2}}
{\beta_{2}-}\frac{\alpha_{2}\gamma_{3}}{\beta_{3}-}\cdots ,
\end{eqnarray}
wherefrom $\nu$ must be determined.
For a specific set of solutions we add
a superscript in each of these quantities.

Moreover, as we are going to explain, solutions with
a phase parameter are also important because they
can be used in two different manners to obtain solutions
for equations having a free constant.

%
%
\subsection*{A. Spheroidal equation}
The sets of solutions are denoted by
by $(U_{i\nu}^{0},\hat{U}_{i\nu}^{\infty},
 U_{i\nu}^{\infty},\widetilde{U}_{i\nu}^{\infty})$
($i=1,2$): the solutions $U_{i\nu}^{0}$
converge for any finite $z$ and the others converge
for $|z|>|z_{0}|$. These solutions were obtained
by integrating the the GSWE \cite{leaver1,eu2,otchik1,mano1},
except $\hat{U}_{i\nu}^{\infty}$ which is derived from
solutions of the general Heun equation
in Appendix B. We write only the first set of solutions
(the second set is obtained from the first by using
the transformation rule $T_{2}$), namely,
\antiletra\letra
\begin{eqnarray}
\begin{array}{l}
U_{1\nu}^{0}(z)=
e^{i\omega z}\displaystyle\sum_{n=-\infty}^{\infty}b_{n}^{(1)}
F\left(\frac{B_{2}}{2}-n-\nu-1,
n+\nu+\frac{B_{2}}{2};B_{2}+\frac{B_{1}}{z_{0}};
\frac{z_{0}-z}{z_{0}}\right),
\vspace{.3cm}\\
\hat{U}_{1\nu}^{\infty}(z)=e^{i\omega z}
\displaystyle\sum_{n=-\infty}^{\infty}
\hat{b}_{n}^{(1)}
\left(\frac{z_{0}-z}{z_{0}} \right)^{-n-\nu-\frac{B_{2}}{2}}\times
\vspace{.3cm}\\
\hspace{1.5cm}\widetilde{F}\left(n+\nu+\frac{B_{2}}{2},n+\nu
+1-\frac{B_{2}}{2}-\frac{B_{1}}{z_{0}};2n+2\nu+2;
\frac{z_{0}}{z_{0}-z}\right),
\vspace{.3cm}\\
U_{1\nu}^{\infty}(z) =e^{i\omega z}z^{1-(B_{2}/2)}
\displaystyle \sum_{n=-\infty}^{\infty}b_{n}^{(1)}
(-2i\omega z)^{n+\nu}
\Psi(n+\nu+1+i\eta,2n+2\nu+2;-2i\omega z),
\vspace{.3cm}\\
\widetilde{U}_{1\nu}^{\infty}(z) =e^{i\omega z}z^{1-(B_{2}/2)}
\displaystyle \sum_{n=-\infty}^{\infty}b_{n}^{(1)}
(2i\omega z)^{n+\nu}
\widetilde{\Phi}(n+\nu+1+i\eta,2n+2\nu+2;-2i\omega z),
\end{array}
\end{eqnarray}
where the functions $\widetilde{F}(a,b;c;y)$ and
$\widetilde{\Phi}(a,b;y)$ are defined by means of
\begin{eqnarray}
\widetilde{F}(a,b;c;y)=\frac{F(a,b;c;y)}{\Gamma(c)},
\ \ \widetilde{\Phi}(a,b;z)=\frac{\Gamma(b-a)}{\Gamma(b)}\Phi(a,b;z)
\end{eqnarray}
and the coefficients $\hat{b}_{n}^{(1)}$ are connected with
$b_{n}^{(1)}$ through
\begin{eqnarray}
\hat{b}_{n}^{(1)}:=(-1)^{n}
\Gamma\left(n+\nu+2-\frac{B_{2
}}{2} \right)
\Gamma\left(n+\nu+1-\frac{B_{1}}{z_{0}}-
\frac{B_{2}}{2} \right)b_{n}^{(1)}, \ \
\mbox{if} \ \ z_{0}\neq 0.
\end{eqnarray}
The coefficients of the recurrence relations (\ref{twosided})
for $b_{n}^{(1)}$ are
\begin{eqnarray}
\begin{array}{l}
\alpha_{n}^{(1)}  =  i\omega z_{0}\frac{\left(n+\nu+2-\frac{B_{2}}{2}\right)
\left(n+\nu+1-\frac{B_{2}}{2}-\frac{B_{1}}{z_{0}}\right)
\left(n+\nu+1-i\eta\right)}
{2(n+\nu+1)\left(n+\nu+\frac{3}{2}\right)},
\vspace{.2cm} \\
\beta_{n}^{(1)} =  -B_{3}-\eta \omega z_{0}-\left(n+\nu+1-\frac{B_{2}}{2}\right)
\left(n+\nu+\frac{B_{2}}{2}\right)
-\frac{\eta \omega z_{0}\left(\frac{B_{2}}{2}-1\right)
\left(\frac{B_{2}}{2}+\frac{B_{1}}{z_{0}}\right)}
{(n+\nu)(n+\nu+1)},
\vspace{.3cm} \\
\gamma_{n}^{(1)} = -i\omega z_{0}\frac{\left(n+\nu+\frac{B_{2}}{2}-1\right)
\left(n+\nu+\frac{B_{2}}{2}+\frac{B_{1}}{z_{0}}\right)
(n+\nu+i\eta)}
{2\left(n+\nu-\frac{1}{2}\right)(n+\nu)}.
\end{array}
\end{eqnarray}
Note that the hypergeometric functions in $U_{1\nu}^{0}$
are well defined only if $B_{2}+(B_{1}/z_{0})\neq 0,-1,-2,\cdots$,
while in $U_{2\nu}^{0}$ we would have the restriction  $B_{2}+(B_{1}/z_{0})\neq 2,3,4,\cdots$.

In these solutions, $\nu$
cannot be integer or half-integer in order
to prevent two dependent  terms
(for different values of $n$) inside the summations.
Suppose, on the contrary, that $2\nu+1=N$
in $U_{1\nu}^{0}$, for some integer $N$.
Then, for $n=n_{1}\ (\forall n_{1})$ and $n=-N-n_{1}$,
the hypergeometric functions in $U_{1\nu}^{0}$ would
be equal on account of their symmetry with respect to
the two first parameters. One may show that this is the same
for the other solutions. In fact, integer or half-integer
values for $\nu$ would lead to infinity coefficients in
the recurrence relations for $b_{n}^{(1)}$.

Now we consider the two possibilities for getting
solutions without phase parameter
from the ones with a phase parameter. The first consists
in breaking off the series by taking $n\geq 0$ in which
case the parameter $\nu$ is given in terms of the
parameters of the differential equation \cite{eu,eu2}.
In particular, the truncation of $U_{1}^{0}$ gives
an expansion in series of Jacobi polynomials which
includes the Baber-Hass\'e solutions (associated Legendre
polynomials) for the angular
equation of the two-center problem of quantum mechanics
\cite{barber} as well as the Fackerell-Crossman solutions for
angular Teukolsky equations \cite{fackerell}.
In this regard, we must do the following correction in
Ref. \cite{eu2}: the truncated expansions in
series of regular confluent hypergeometric
functions converge for $|z|>|z_{0}|$, rather than for any
$z$ as stated there.  Therefore, such solutions are not suitable
to solve an angular equation because in this
case $|z|\leq|z_{0}|$.

The other possibility gives solutions in terms of
doubly infinite series ($-\infty<n<\infty$). These are obtained
by ascribing any convenient value to $\nu$, excepting
integer or half-integer values.
Such solutions exhibit properties that may be quite different
from the properties of  the one-sided series ($n\geq 0$),
as illustrated by the following solution for WHE which
we derive from a solution for the GSWE. Proceeding as in Sec. II B
and using the relation \cite{abramowitz}
\antiletra
\begin{eqnarray}
F(-a,a; 1/2;\sin^2u)=\cos(2au),
\end{eqnarray}
we find that the solution $U_{1\nu}^{0}(z)$ affords an even
solution for the WHE, namely,
\begin{eqnarray}
W_{1\nu}^{0}(u)=e^{(\xi/2) \cos^2(\kappa u)} \displaystyle \sum_{n=-\infty}^{\infty}b_{n}
\cos[(2n+2\nu+1)\kappa u],
\end{eqnarray}
where the recurrence relations for $b_{n}$  are ($-\infty<n<\infty$)
\begin{eqnarray}
 \xi\left(n+\nu+\frac{1}{2}-\frac{p}{2}\right)
b_{n+1}+\left[  \vartheta -\left(2n+2\nu+1\right)^{2}\right] b_{n}
- \xi \left(n+\nu+\frac{1}{2}+\frac{p}{2}\right)b_{n-1}=0.
\end{eqnarray}
Now, if there is a free parameter
in the differential equation, we can choose any admissible
value for $\nu$ and, in particular,
 \begin{eqnarray}
2\nu+1=l/m, \ \ \kappa=1,
\end{eqnarray}
where $l$ and $m$ are integers prime to one another,
so that $1<l<m$. Then, we find the following two-sided Ince
solution, $W_{1}^{I}(u)$, for the WHE \cite{ince}
 \begin{eqnarray}
W_{1}^{I}(u)=e^{\frac{\xi}{2} \cos^2(u)}
\displaystyle \sum_{n=-\infty}^{\infty}b_{n}^{(1)}\cos
\left[ \left( 2n+\frac{l}{m}\right)  u\right],
\end{eqnarray}
having period $2\pi m$, $m>1$,
in opposition to the solutions obtained from the truncated
solutions \cite{eu2} (period $\pi$ or $2\pi$).  An odd solution
may also be found from the $U_{2\nu}^{0}(z)$
of the (omitted) second set of solutions.

Although this example is concerned only with the periodicity
and parity of solutions for the Whittaker-Hill equation, the analysis
can be extended to other aspects of the GSWE and their
limiting cases. Next we use two-sided solutions similar to
the above one to complete
the solutions of the Schr\"odinger equation considered in section III.

\subsection*{B. Double-confluent Heun and Schr\"odinger
equations}
From the previous solutions there is no
difficulty in finding solutions for the DCHE by means of the
Leaver limit, provided
that we use the formulas \cite{erdelyi1}
\begin{eqnarray}\label{limites}
\begin{array}{l}
\displaystyle\lim_{c\rightarrow \infty}
F\left(a,b;c;1-\frac{c}{y}\right)=
y^a\Psi(a,a+1-b;y),\vspace{3mm}\\
\displaystyle \lim_{b\rightarrow \infty}F\left(a,b;c;\frac{y}{b}\right)=
\Phi(a,c;y).
\end{array}
\end{eqnarray}
We find the Leaver solutions \cite{leaver1,eu2}
\begin{eqnarray}
\begin{array}{l}
U_{1\nu}^{0}(z)  =e^{i\omega z}z^{-\frac{B_{2}}{2}}\displaystyle \sum_{n=-\infty}^{\infty}
b_{n}^{(1)}
\left(\frac{B_{1}}{z}\right)^{n+\nu}
\Psi\left(n+\nu+\frac{B_{2}}{2},2n+2\nu+2;
\frac{B_{1}}{z}\right),
\vspace{0.3cm}\\
\hat{U}_{1\nu}^{\infty}(z)  =e^{i\omega z}z^{-\frac{B_{2}}{2}}\displaystyle \sum_{n=-\infty}^{\infty}
b_{n}^{(1)}
\left(-\frac{B_{1}}{z}\right)^{n+\nu}
\tilde{\Phi}\left(n+\nu+\frac{B_{2}}{2},2n+2\nu+2;
\frac{B_{1}}{z}\right),
\vspace{0.3cm}\\
U_{1\nu}^{\infty}(z) =e^{i\omega z}z^{1-\frac{B_{2}}{2}}
\displaystyle \sum_{n=-\infty}^{\infty}b_{n}^{(1)}(-2i\omega z)^{n+\nu}
\Psi(n+\nu+1+
i\eta,2n+2\nu+2;-2i\omega z),
\vspace{3mm}\\
\widetilde{U}_{1\nu}^{\infty}(z) =e^{i\omega z}z^{1-\frac{B_{2}}{2}}
\displaystyle \sum_{n=-\infty}^{\infty}b_{n}^{(1)}
(2i\omega z)^{n+\nu}
\tilde{\Phi}(n+\nu+1+i\eta,2n+2\nu+2;-2i\omega z),
\end{array}
\end{eqnarray}
with the following coefficients in the recurrence relations (\ref{twosided})
for $b_{n}^{(1)}$:
\begin{eqnarray}
\begin{array}{l}
\alpha_{n}^{(1)}  =  i\omega B_{1}\frac{
\left(n+\nu+2-\frac{B_{2}}{2}\right)(n+\nu+1-i\eta)}
{2(n+\nu+1)\left(n+\nu+\frac{3}{2}\right)},
\vspace{3mm}\\
\beta_{n}^{(1)}  = B_{3}-\frac{1}{4}
\left( B_{2}-1\right)^2 +\left(n+\nu+\frac{1}{2}\right)^2+
\frac{\eta \omega B_{1}\left(\frac{B_{2}}{2}-1\right)}
{(n+\nu)(n+\nu+1)},
\vspace{3mm}\\
\gamma_{n}^{(1)}=
i\omega B_{1} \frac{\left(n+\nu+\frac{B_{2}}{2}-1\right)(n+\nu+
i\eta)}{2(n+\nu)\left(n+\nu-\frac{1}{2}\right)}.
\end{array}
\end{eqnarray}
These four solutions result as the Leaver limits
of solutions for the GSWE because in Sec. VI.A we
have chosen a set formed by four solutions. They
can also be obtained by starting with the two
expansions in series of confluent hypergeometric
functions for the GSWE and, then, using a symmetry
of the DCHE \cite{leaver1,eu2}.

Now we use these solutions of the DCHE to find
infinite-series solutions
for the quasi-exactly solvable double-Morse
potential discussed in section III.
For this is necessary to match two solutions
having different regions of convergence, a procedure
which has been employed to
find solutions for the radial Teukolsky equations in Kerr backgrounds
\cite{otchik1,mano1,mano2,mano3}. In this manner, the
missing portion of the energy spectrum may be obtained.

By inserting the previous $U_{1\nu}^{0}$ and
$U_{1\nu}^{\infty}$ into
Eq. (\ref{sol}) and using the parameters
given in Eq. (\ref{paramzas}),
we get, respectively,
\begin{eqnarray}
\begin{array}{l}
\phi_{1\nu}^{0}(z) =
\exp\left[-\frac{B}{4}\left(z+\frac{1}{z}\right)\right]
\displaystyle \sum_{n= -\infty}^{\infty}b_{n}^{(1)}
\left(\frac{B}{2z}\right)^{n+\nu+\frac{1}{2}}\Psi
\left(n+\nu+\frac{1}{2}+\frac{C}{2}-s,2n+2\nu+2;\frac{B}{2z}\right),
\vspace{3mm}\\
\phi_{1\nu}^{\infty}(z) =
\exp\left[-\frac{B}{4}\left(z+\frac{1}{z}\right)\right]
\displaystyle \sum_{n= -\infty}^{\infty}b_{n}^{(1)}
\left(\frac{Bz}{2}\right)^{n+\nu+\frac{1}{2}}\Psi\left(n+\nu+\frac{1}{2}
-\frac{C}{2}-s,2n+2\nu+2;
\frac{Bz}{2}\right),
\end{array}
\end{eqnarray}
where in the recurrence relations
(\ref{twosided}) for $b_{n}^{(1)}$ now we have ($B>0$, $C>0$)
\begin{eqnarray}
\begin{array}{l}
\alpha_{n}^{(1)}  = -\frac{B^2}{16} \frac{\left(n+\nu+\frac{3}{2}-\frac{C}{2}+s\right)
\left(n+\nu+\frac{3}{2}+\frac{C}{2}+s\right)}
{\left(n+\nu+1\right)\left(n+\nu+\frac{3}{2}\right)} ,
\vspace{.3cm} \\
\beta_{n}^{(1)}  = {\cal E}+\frac{B^2}{8}-\frac{C^2}{4}
+\left(n+\nu+\frac{1}{2}\right)^2
-\frac{B^2}{32}\frac{\left[C^2-\left(1+2s\right)^2\right]}
{\left(n+\nu\right)\left(n+\nu+1\right)},
\vspace{.3cm} \\
\gamma_{n}^{(1)}= -\frac{B^2}{16}\frac{\left(n+\nu+\frac{C}{2}-\frac{1}{2}-s\right)
\left(n+\nu-\frac{C}{2}-\frac{1}{2}-s\right)}{\left(n+\nu\right)
\left(n+\nu-\frac{1}{2}\right)}.
\end{array}
\end{eqnarray}
For infinite series, these solutions satisfy the regularity
conditions (\ref{regularity}) since we have ($B>0$)
\begin{eqnarray*}
\begin{array}{l}
\displaystyle \lim_{ z\rightarrow 0}\phi_{1\nu}^{0}(z)
\sim  z^{\frac{C}{2}-s}
\exp\left[-\frac{B}{4}\left(z+\frac{1}{z}\right)\right]\to 0,
\vspace{3mm}\\
\displaystyle \lim_{ z\rightarrow \infty}\phi_{1\nu}^{\infty}(z)
\sim z^{\frac{C}{2}+s}
\exp\left[-\frac{B}{4}\left(z+\frac{1}{z}\right)\right] \to 0,
\end{array}
\end{eqnarray*}
where we have used the relation (\ref{asymptotic}).
However, in order to guarantee that in fact these
are solutions given by infinite
series, we must choose suitable values for the parameter $\nu$.
Recalling that $\nu$ cannot be integer
or half-integer and that $s$ is a non-negative integer or half-integer,
we find that there are three different
cases to be considered.

{\it First case:} $C$ is integer or half-integer. In this case
we can take $\nu=1/3$ to find
\begin{eqnarray}
\begin{array}{l}
\alpha_{n}^{(1)}=-\frac{B^2}{16} \frac{\left(n+s+\frac{3}{2}+\frac{1}{3}-\frac{C}{2}\right)
\left(n+s+\frac{3}{2}+\frac{1}{3}+\frac{C}{2}\right)}
{\left(n+1+\frac{1}{3}\right)\left(n+\frac{1}{3}+\frac{3}{2}\right)},
\vspace{3mm}\\
\beta_{n}^{(1)}  = {\cal E}+\frac{B^2}{8}-\frac{C^2}{4}
+\left(n+\frac{1}{3}+\frac{1}{2}\right)^2
-\frac{B^2}{32}\frac{\left[C^2-\left(1+2s\right)^2\right]}
{\left(n+\frac{1}{3}\right)\left(n+1+\frac{1}{3}\right)},
\vspace{.3cm} \\
\gamma_{n}^{(1)}= -\frac{B^2}{16}\frac{\left(n-s-\frac{1}{2}+\frac{1}{3}+
\frac{C}{2}\right)
\left(n-s-\frac{1}{2}+\frac{1}{3}-\frac{C}{2}\right)}
{\left(n+\frac{1}{3}\right)
\left(n+\frac{1}{3}-\frac{1}{2}\right)}.
\end{array}
\end{eqnarray}
Thence, the numerators of $\alpha_{n}^{(1)}$
and $\gamma_{n}^{(1)}$ do not vanish since $(1/3)+(C/2)$ and $(1/3)-(C/2)$ are not integer or half-integer. Therefore, we have
a pair of solutions given by two-sided infinite series.

{\it Second case:} $C$ is not integer or half-integer,
but $s$ is integer. In this case we can take
\begin{eqnarray*}
\nu=\frac{C}{2}+1
\end{eqnarray*}
which is different of integer or half-integer.
Then,
\begin{eqnarray}
\begin{array}{l}
\alpha_{n}^{(1)}  = -\frac{B^2}{16} \frac{\left(n+s+\frac{5}{2}\right)
\left(n+s+\frac{5}{2}+C\right)}
{\left(n+2+\frac{C}{2}\right)\left(n+\frac{5}{2}+\frac{C}{2}\right)} ,
\vspace{.3cm} \\
\beta_{n}^{(1)}  = {\cal E}+\frac{B^2}{8}-\frac{C^2}{4}
+\left(n+\frac{3}{2}+\frac{C}{2}\right)^2
-\frac{B^2}{32}\frac{\left[C^2-\left(1+2s\right)^2\right]}
{\left(n+1+\frac{C}{2}\right)\left(n+2+\frac{C}{2}\right)},
\vspace{.3cm} \\
\gamma_{n}^{(1)}=
 -\frac{B^2}{16}\frac{\left(n-s+\frac{1}{2}+C\right)
\left(n-s+\frac{1}{2}\right)}{\left(n+1+\frac{C}{2}\right)
\left(n+\frac{1}{2}+\frac{C}{2}\right)}.
\end{array}
\end{eqnarray}
Thus the numerators of $\alpha_{n}^{(1)}$
and $\gamma_{n}^{(1)}$ cannot vanish and, once more,
we have two-sided infinite series.

{\it Third case:} $C$ is not integer or half-integer, but $s$ is half-integer.
In this case we can take
\begin{eqnarray*}
\nu=\frac{C}{2}+\frac{1}{2}
\end{eqnarray*}
which again imply that the numerators of $\alpha_{n}^{(1)}$
and $\gamma_{n}^{(1)}$ cannot vanish since
\begin{eqnarray}
\begin{array}{l}
\alpha_{n}^{(1)}  = -\frac{B^2}{16} \frac{\left(n+s+2\right)
\left(n+s+2+C\right)}
{\left(n+\frac{3}{2}+\frac{C}{2}\right)\left(n+2+\frac{C}{2}\right)} ,
\vspace{.3cm} \\
\beta_{n}^{(1)}  = {\cal E}+\frac{B^2}{8}-\frac{C^2}{4}
+\left(n+1+\frac{C}{2}\right)^2
-\frac{B^2}{32}\frac{\left[C^2-\left(1+2s\right)^2\right]}
{\left(n+\frac{1}{2}+\frac{C}{2}\right)\left(n+\frac{3}{2}
+\frac{C}{2}\right)},
\vspace{.3cm} \\
\gamma_{n}^{(1)}= -\frac{B^2}{16}\frac{\left(n-s+C\right)
\left(n-s\right)}{\left(n+\frac{1}{2}+\frac{C}{2}\right)
\left(n+\frac{C}{2}\right)}.
\end{array}
\end{eqnarray}

Therefore, in principle this problem can be solved
by matching the two
double-sided solutions which converge over different
ranges of the independent
variable $z$. The energy spectrum can be obtained
by solving the characteristic
equation (\ref{twosidedcharac}). Then, for a given
value of $s$, besides the energy levels provided
by the quasi-polynomials solutions of Sec. III.A,
there is the spectrum resulting from the present
infinite-series solutions. In addition, for $C=0$,
the first case affords solutions in terms of doubly infinite series
for the symmetric double-Morse potential (WHE).

%
%
\section*{VII. Conclusions}
\indent
We have seen that the Leaver and Whittaker-Ince limits of the
generalized spheroidal wave equation (GSWE) allow us
to generate solutions for three other equations with
different types of singular points.  We have also obtained
new solutions for the Mathieu and Whittaker-Hill equations
by regarding these as particular cases of the GSWE and
its limiting cases. As an application we have considered the
Schr\"odinger equation with an asymmetric double-Morse
potential.

First we have constructed a set of solutions to the GSWE
in such a manner that the three solutions of
each set converge over different
regions but have series coefficients proportional to one
another. Actually, this is an extention of an approach already
used in a previous work \cite{eu2} (revisited in Sec. VI),
the difference being that in Sec. II the
initial set of solutions is given by a series of ascending
powers of $z-z_{0}$,
due to Barber and Hass\'e, and two expansions in series of confluent
hypergeometric functions. From the initial set other ones have been
generated by means of transformation rules which reflect
the symmetries of the GSWE. The importance of these rules
can be appreciated by considering the properties of each set of
solutions that we have found for the Mathieu and Whittaker-Hill equations.

From solutions of the GSWE, sets of solutions for the
three limiting cases have been found following the
diagram (\ref{scheme}).
Thus, the solutions for the double-confluent Heun equation
(DCHE) have resulted by means of the Leaver limit.
On the other hand, in the Whittaker-Ince limit for the GSWE and DCHE,
the expansions in series of confluent hypergeometric
functions have afforded expansions in series of Bessel
functions as solutions for Eqs. (\ref{incegswe})
and (\ref{incedche}).

In this process for generating solutions,
the Barber-Hass\'e expansions have led to
some known solutions for the DCHE, Mathieu and
Whittaker-Hill equations, but these solutions are
accompanied by two new solutions belonging to each set.
Besides this, the Whittaker-Ince limit of the
Barber-Hass\'e expansions has also given solutions which
include as particular case the solution already proposed by
Malmendier \cite{malmendier} to solve a wave equation
which arises from separation of
variables of the Laplace-Beltrami operator for a scalar field
on the Eguchi-Hanson space. However, now we have found
two expansions in series of Bessel
functions, both of them convergent in the vicinity of
$z=\infty$ and satisfying three-term recurrence relations
for the series coefficients, in contrast with the
solution proposed by Malmendier (four-term recurrence relations).

The properties of the solutions for the  Mathieu and
Whittaker-Hill equations depend on whether
these solutions come from the GSWE
or DCHE. We have seen that the unique solutions
for the WHE which are even or odd and have period
$\pi$ or $2\pi$ result from the solutions for the GSWE (Sec. II B).
The Whittaker-Ince limit transfer these properties
to the solutions of the Mathieu equation
given in Sec. IV B, each set including
one of the Lindemann-Stieltjes solutions and two new
solutions in series of Bessel functions. The
latter solutions demand further investigation
since they are different from the usual expansions
in series of Bessel functions.

On the other hand, the solutions for WHE
considered as a DCHE (Sec. III B)
do not possess such properties of parity and periodicity
(even or odd, period $\pi$ or $2\pi$).
The same is true for the solutions of the Mathieu
equation when this last is treated as a particular case
(not as Whittaker-Ince limit) of the WHE (Sec. II C) or as
a particular case of the DCHE (Sec. III C). However, solutions
for the Mathieu and Whittaker-Hill equations which do not
present the usual properties of parity and periodicity
may be important to solve initial value problems instead of
boundary value problems \cite{McLachlan}.


Up to section V we have assumed that there is a
free constant in the differential equation,
and all the solutions are one-sided in the sense that
the summation index $n$ runs from $0$ to $\infty$.
In this case, the free constant is determined from a
characteristic  equation
which must be fulfilled in order to assure the series
convergence. However, in section VI we have reconsidered
a set of solutions in series of hypergeometric and confluent
hypergeometric functions which converge when there
is no free  constant in the differential equation.
In such solutions there is a parameter $\nu$, called
phase or characteristic parameter \cite{leaver1,arscott},
which must be adjusted so that a characteristic equation
is satisfied in order to guarantee the series convergence.

The solutions with the phase parameter $\nu$ are
two-sided ($-\infty<n<\infty$) and also admit
both the Leaver and Whittaker-Ince limits. In addition,
they allow us to find one-sided and two-sided
solutions for an equation having a free parameter.
The one-sided solutions are obtained by the restriction
$n\geq 0$ in which case $\nu$ becomes determined in terms
of other parameters of the equation \cite{eu2}.
The two-sided solutions are generated by choosing
some of the admissible values for $\nu$, that is,
some $\nu$ different from integer or half-integer.
These two-sided solutions have been used in section
VI B to find infinite-series solutions for the
Schr\"odinger equation with an asymmetric
double-Morse potential, by reducing that equation
to a DCHE.


The asymmetric double-Morse potential
considered here is a quasi-exactly solvable (QES) potential
for which a part of the energy
spectrum can be determined  from the recurrence relations
for the coefficients of finite-series eigensolutions,
as the ones found in Sec II A.  On the other hand, in Sec.
VI B we have shown that, for a given value of the the parameter $s$,
the other portion of the spectrum can be derived as solutions
for a characteristic equation associated with eigenfunctions
given  by infinite two-sided series, provided that we
accept to match two solutions which converge over different
ranges of the independent variable.
When the parameter $C$ vanishes, the asymmetric
potential degenerate to a symmetric one,
whereas the Schr\"odinger equation degenerate to
a WHE which has already been solved as a GSWE \cite{eu2}.
Therefore, this symmetric case is an instance of a
WHE which can be regarded both as a DCHE and a GSWE.

On the other hand, for the trigonometric and hyperbolic
QES potentials given by Usheveridze \cite{usheveridze1},
the Schr\"odinger equation can be transformed into
GSWEs \cite{lemieux}. Thus, the solutions
for the GSWE given in this paper may be used to verify if
it is possible to find the whole energy spectra for these potentials,
as in the case of the double-Morse potentials.

Another possible application consists in taking the solutions
for the GSWE as a guide to find new solutions for the
(general) Heun equation, since the  former equation
is a confluent case of the latter. This would
complement the procedure presented in the preceding
sections, because eventually it would add
the connection: general Heun equation $\to$ GSWE into
the diagram (\ref{scheme}). In fact, this link is required
by the study of Teukolsky equations for gravitational backgrounds
of black holes, as we have mentioned in the first section.
Although the Heun equation is not subject of the present paper,
in Appendix B we have introduced some preliminary
results which indicate how the above goal could be attained.

\section*{Acknowledgments}
\indent
I am indebted to Dr. L\'ea Jaccoud El-Jaick for her
collaboration concerning results of section VI.B and
Appendix B. I also thank Dr. Herman J. M. Cuesta
for his reading of the manuscript and fruitful criticisms.

%
\section*{Appendix A: Solutions in series of Bessel functions}
\protect\label{A}
\setcounter{equation}{0}
\renewcommand{\theequation}{A\arabic{equation}}
\indent
To find directly the solution $U_{1\nu}^{\infty}(z)$ for the
Whittaker-Ince limit
of the generalized spheroidal wave equation,
\begin{eqnarray}\label{B0}
z(z-z_{0})\frac{d^{2}U}{dz^{2}}+(B_{1}+B_{2}z)
\frac{dU}{dz}+
\left[B_{3}+q(z-z_{0})\right]U=0,\ (q\neq0),
\end{eqnarray}
we perform the substitutions
\begin{eqnarray}\label{B1}
t=2i\sqrt{qz},\ \ U_{1\nu}^{\infty}(z)=t^{1-B_{2}}Y(t),
\end{eqnarray}
to obtain
\begin{eqnarray*}
t^2\frac{d^2Y}{dt^2}+t\frac{dY}{dt}-
t^2Y=-4qz_{0}\frac{d^2Y}{dt^2}-
\frac{4q(z_{0}-2B_{1}-2B_{2}z_{0})}{t}
\frac{dY}{dt} \nonumber\\
+\left[4q(1-B_{2})\frac{2B_{1}+B_{2}z_{0}+z_{0}}
{t^2}+(1-B_{2})^2+4qz_{0}-4B_{3}\right]Y.
\end{eqnarray*}
Next, by expanding $Y(t)$ in series of
modified Bessel functions of the second
kind  $K_{\lambda}(t)$ according to
\begin{eqnarray}\label{B2}
Y(t)= \sum_{n=0}^{\infty}d_{n}
t^{-n}K_{\lambda}(t), \mbox{where} \ \lambda=n+B_{2}-1,
\end{eqnarray}
we find
\begin{eqnarray}\label{B3}
\frac{dY}{dt}= \sum_{n=0}^{\infty}d_{n}
t^{-n}\left[\frac{dK_{\lambda}}{dt}-
\frac{n}{t}K_{\lambda}\right], \
\frac{d^{2}Y}{dt^{2}}= \sum_{n=0}^{\infty}d_{n}
t^{-n}\left[\frac{d^{2}K_{\lambda}}{dt^{2}}- \frac{2n}{t}\frac{dK_{\lambda}}{dt}+\frac{n(n+1)}{t^{2}}
K_{\lambda}\right].
\end{eqnarray}
Inserting (\ref{B2}) and  (\ref{B3}) into the differential
equation for $Y(t)$ and
using the modified Bessel equation,
\begin{eqnarray*}
t^2\frac{d^2K_{\lambda}}{dt^2}+t\frac{dK_{\lambda}}{dt}-
t^2K_{\lambda}=\lambda^{2}K_{\lambda},
\end{eqnarray*}
it results
\begin{eqnarray}\label{B4}
&&\displaystyle \sum_{n=0}^{\infty}
d_{n}t^{-n}\left[\lambda^2+n^2+4B_{3}-4qz_{0}
-(1-B_{2})^2\right]
K_{\lambda}=\nonumber\\
&&\displaystyle \sum_{n=0}^{\infty}d_{n}t^{-n}
\left[-4qz_{0}\frac{d^2K_{\lambda}}{dt^2}+
2nt\frac{dK_{\lambda}}{dt}+4q\left(2nz_{0}-z_{0}+
2z_{0}B_{2}+2B_{1}\right)
\frac{1}{t}\frac{dK_{\lambda}}{dt}\right]
\nonumber\\
&&-4q\displaystyle \sum_{n=0}^{\infty}
d_{n}t^{-n-2}\left[n(z_{0}n+2z_{0}+2B_{1})-
(1-B_{2})(z_{0}B_{2}+2B_{1}+z_{0})\right]
K_{\lambda}.
\end{eqnarray}
From the difference-differential relations for
$K_{\lambda}$ \cite{luke}, we find that
\begin{eqnarray*}
&&\frac{d^2K_{\lambda}}{dt^2}=
K_{\lambda}+\lambda(\lambda-1)t^{-2}K_{\lambda}+
t^{-1}K_{\lambda+1},\\
&&t\frac{dK_{\lambda}}{dt}=
-\lambda K_{\lambda}-
t K_{\lambda-1},\\
&&\frac{1}{t}\frac{dK_{\lambda}}{dt}=
\lambda t^{-2} K_{\lambda}-
t^{-1}K_{\lambda+1}.
\end{eqnarray*}
Thence, taking into account that $\lambda=n +B_{2}-1$,
Eq. (\ref{B4}) simplifies to
\begin{eqnarray*}\begin{array}{l}
\displaystyle \sum_{n=1}^{\infty}
\frac{n}{2}d_{n}
t^{-n+1}K_{n+B_{2}-2}+
\displaystyle \sum_{n=0}^{\infty}
[n(n+B_{2}-1)+B_{3}]d_{n}
t^{-n}K_{n+B_{2}-1}+\vspace{3mm}\\
\displaystyle \sum_{n=0}^{\infty}
\left[2z_{0}q\left(n+B_{2}+\frac{B_{1}}{z_{0}}\right)\right]d_{n}
t^{-n-1}K_{n+B_{2}}=0,
\end{array}\end{eqnarray*}
since the coefficient of $t^{-n-2}K_{\lambda}$
vanishes. By the replacements
$n\rightarrow m+1$ and
$n\rightarrow m-1$ in the first and third
terms, respectively, this equation becomes
\begin{eqnarray*}\begin{array}{l}
\left[ \frac{1}{2}d_{1}+B_{3}d_{0}\right] K_{B_{2}-1}+
\displaystyle \sum_{n=1}^{\infty}\left\{
\frac{1}{2}(n+1)d_{n+1}+\right.
[n(n+B_{2}-1)+B_{3}]d_{n}+
\vspace{3mm}\\
\left.
2z_{0}q\left(n+B_{2}+\frac{B_{1}}{z_{0}}-1\right)d_{n-1}\right\}
t^{-n}K_{n+B_{2}-1}=0.
\end{array}\end{eqnarray*}
Requiring that the coefficient of each independent
term of this equation to vanish,
we get the relations ($d_{-1}=0$)
\begin{eqnarray}\label{B5}
\frac{1}{2}(n+1)d_{n+1}+
[n(n+B_{2}-1)+B_{3}]d_{n}+
2z_{0}q\left(n+B_{2}+\frac{B_{1}}{z_{0}}-1\right)d_{n-1}=0,\ n\geq 0.
\end{eqnarray}
Then, from Eqs. (\ref{B1}) and (\ref{B2}) we have
\begin{eqnarray}\label{B6}
U_{1\nu}^{\infty}(z(t))=t^{1-B_{2}}
\displaystyle\sum_{n=0}^{\infty}d_{n}
t^{-n}K_{n+B_{2}-1,}(t)\propto\
z^{(1-B_{2})/2}\displaystyle\sum_{n=0}^{\infty}2^{-n}d_{n}
(i\sqrt{qz})^{-n}K_{n+B_{2}-1,}(2i\sqrt{qz}).
\end{eqnarray}
By putting $c_{n}^{(1)}=2^{-n}d_{n}$ in
Eqs. (\ref{B5}) and (\ref{B6}), we recover
the solution given in Sec. IV. The derivation of the
expansion in series of $J_{\lambda}(2\sqrt{qz})$
is similar to the previous one. Thus, the Whittaker-Ince limits
found in Sec. IV in fact satisfy the differential equation (\ref{B0}).

\section*{Appendix B: General and confluent Heun equations}
\protect\label{B}
\setcounter{equation}{0}
\renewcommand{\theequation}{B\arabic{equation}}
The relationship between solutions of general Heun equation
and GSWE (confluent Heun) is interesting
by itself and becomes, in the context
of relativistic astrophysics, important in the study
of the Teukolsky equations of the gravitational backgrounds
of black holes. In effect, for the Kerr metric these
equations turn out to be GSWEs which afford DCHEs
in the extreme upper limit of the rotation parameter
\cite{leaver1}. On the other hand,  the
Teukolsky equations for the Kerr-de Sitter metric are (general)
Heun equations which, in turn, reduce to GSWEs
when the cosmological constant $\Lambda$
vanishes \cite{mano4,mano5}. Therefore, in such problems
we have the connection:
Heun equation $\to$ GSWE $\to$ DCHE, and this
requires a generalization of the diagram (\ref{scheme}).

In the following we exhibit two  Erd\'elyi's solutions
for the Heun equation
\cite{erdelyi2} which lead to the expansions in
hypergeometric functions given Sec. VI for the
GSWE. From these Erd\'elyi solutions we
explain how to obtain two additional solutions
which yield the expansions in series of confluent
hypergeometric functions for the GSWE as well.

To begin with, let there be the Heun equation
\cite{ronveaux,heun,maier}
\begin{eqnarray}\label{heun}
\frac{d^{2}H}{dx^{2}}+\left(\frac{\gamma}{x}+
\frac{\delta}{x-1}+\frac{\epsilon}{x-a}\right)\frac{dH}{dx}+
\frac{\alpha \beta x-q}{x(x-1)(x-a)}H=0, \ \ \epsilon=\alpha+\beta+1-\gamma-\delta,
\end{eqnarray}
where $a\in \mathbb{C}\setminus\{0,1\}$
and $x=0,1,a,\infty$ are regular singular points with indicial
exponents given by $\{0,1-\gamma\}$, $
\{0,1-\delta\}$, $\{0,1-\epsilon\}$ and $\{\alpha, \beta\}$,
respectively. For $a=0$ and $a=1$ the equation
can be reduced to the hypergeometric equation
by changes of variables; also for $\epsilon=0$ and
$q=a\alpha\beta$ it degenerates to
the hypergeometric equation. There is no
possibility of confusing the auxiliary parameter
$q$ with the parameter $q$ of the Whittaker-Ince limits
given in Eq. (\ref{ince}) neither the singularity
parameter $a$ with the constant $a$ of the Mathieu
equation (\ref{mathieu}) because the $q$  and $a$
of the Heun equation disappear in the confluent limit of
the equation (see Eq. (\ref{confluence})).

As in Ref. \cite{mano5}, we denote by
$H_{1\nu}^{\{0,1\}}(x)$
and $\hat{H}_{1\nu}^{\{a,\infty\}}(x)$ the two
Erd\'elyi solutions
in series of hypergeometric functions, where the superscripts
mean that the first solution converges in a
domain containing the singular points $0$ and $1$, and the
second, in a domain containing $a$ and $\infty$. These
solutions are \cite{erdelyi2}
(once more $\nu$ is different from integer or half-integer)
\begin{eqnarray}\begin{array}{l}
H_{1\nu}^{\{0,1\}}(x)=\displaystyle\sum_{n=-\infty}^{\infty}
b_{n}^{(1)}F\left( -n-\nu-1+\frac{\gamma+\delta}{2},n+\nu+\frac{\gamma+\delta}{2};
\gamma;x\right) ,
\vspace{3mm}\\
\hat{H}_{1\nu}^{\{a,\infty\}}(x)=
\displaystyle\sum_{n=-\infty}^{\infty}
\hat{b}_{n}^{(1)}
x^{-n-\nu-(\gamma+\delta)/2}\widetilde{F}\left(n+\nu
+1+\frac{\delta-\gamma}{2},n+\nu+\frac{\gamma+\delta}{2};
2n+2\nu+2;
\frac{1}{x}\right),
\end{array}
\end{eqnarray}
where
\begin{eqnarray}
\hat{b}_{n}^{(1)}=(-1)^{n}
\Gamma\left(n+\nu+2-\frac{\gamma+\delta}{2} \right)
\Gamma\left(n+\nu+
1+\frac{\delta-\gamma}{2} \right)b_{n}^{(1)}.
\end{eqnarray}
In the recurrence relations (\ref{twosided})
for $b_{n}^{(1)}$ the coefficients are
\begin{eqnarray}\label{coef}
&&\alpha_{n}^{(1)}=
-\frac{\left( n+\nu+2-\frac{\gamma+\delta}{2}\right) \left( n+\nu+1-\alpha+\frac{\gamma+\delta}{2}\right)
\left( n+\nu+1-\beta+\frac{\gamma+\delta}{2}\right)
\left( n+\nu+1-\frac{\gamma-\delta}{2}\right) }
{4\left(n+\nu+1\right)
\left(n+\nu+\frac{3}{2}\right)},
\nonumber\\
\nonumber\\
&&\beta_{n}^{(1)}=\left(\frac{1}{2} -a\right)
\left( n+\nu\right)
\left( n+\nu+1\right)+\frac{1}{4}\left[2\alpha\beta+
(\alpha+\beta)(\gamma-\delta)+(\gamma+\delta)(1-\gamma) \right]
\nonumber\\
\nonumber\\
&&\hspace{1cm}
-\frac{1}{4}a(\gamma+\delta)(2-\gamma-\delta) -q+
\frac{(\gamma-\delta)(\gamma+\delta-2)(2\alpha-\gamma-\delta)
(2\beta-\gamma-\delta)}
{32(n+\nu)(n+\nu+1)},
\nonumber\\
\nonumber\\
&&\gamma_{n}^{(1)}=-\frac{\left( n+\nu+\alpha-
\frac{\gamma+\delta}{2}\right) \left( n+\nu+\beta-
\frac{\gamma+\delta}{2}\right) \left( n+\nu-1+\frac{\gamma+\delta}{2}\right)
\left( n+\nu+\frac{\gamma-\delta}{2}\right) }
{4\left(n+\nu-\frac{1}{2}\right) \left(n+\nu\right) }.
\end{eqnarray}

In these solutions the parameter $\nu$
which appears in some other works, for example in Ref. \cite{ronveaux}, have been replaced by the
expression on right-hand side of the relation
\begin{eqnarray}\label{susuki}
\nu\to \nu+1-\frac{\gamma+\delta}{2}.
\end{eqnarray}
This redefinition is due to Suzuki, Takasugi
and  Umetsu \cite{mano5}. It is useful to get the
limits for the GSWE, as well as to obtain new solutions
from the preceding ones by means of
transformations like that written in
Eq. (\ref{maier105}), simply because it is not
necessary to transform $\nu$ in order to get
solutions having the same series coefficients.

The above solutions are valid only if $a\notin[0,1]$
\cite{erdelyi2}. Since
this restriction comes from the analysis of the recurrence
relations for the coefficients $b_{n}^{(1)}$  as $n\to \pm\infty$,
it is transferred to solutions whose coefficients are proportional
to $b_{n}^{(1)}$. The solution
$H_{1\nu}^{\{0,1\}}(x)$ converges in the interior of an ellipse
having foci at $0$ and $1$ and passing through the point $a$,
provided that the characteristic equation is satisfied; the solution
$\hat{H}_{1\nu}^{\{a,\infty\}}(x)$ converges in the entire complex
plane excepting the line joining the points $0$ to
$1$ \cite{erdelyi2}. Therefore,
we are dealing with a pair of solutions convergent over
different domains.


To derive the GSWE as a confluence of
Heun's equation (\ref{heun}), first we rewrite the latter as
\begin{eqnarray*}
x(x-1)\left(\frac{x}{a} -1\right) \frac{d^2 H}{dx^2}+
\left[\gamma(x-1)\left( \frac{x}{a}-1\right) +
\delta x \left( \frac{x}{a}-1\right) +\frac{\epsilon}{a} x(x-1)\right]
\frac{dH}{dx}+\left[\alpha\frac{\beta}{a}x-\frac{q}{a}\right]H =0.
\end{eqnarray*}
Then, by letting that
\begin{eqnarray}\label{confluence}
a, \ \beta,\ q\to\infty\ \mbox{such that}\ \frac{\beta}{a} \to
\frac{\epsilon}{a}\to -\rho, \ \frac{q}{a}\to -\sigma, \ H(x)\to h(x),
\end{eqnarray}
$\rho$ and $\sigma$ being constants, we find the GSWE
\cite{ronveaux}
\begin{eqnarray}\label{arscott}
x(x-1)\frac{d^2 h}{dx^2}+[-\gamma+(\gamma+\delta)x+\rho x(x-1)]
\frac{dh}{dx}+[\alpha\rho x-\sigma]h=0,
\end{eqnarray}
which, however, is not in the form suitable to get the
DCHE by means of the Leaver limit. Despite this, from
the previous solutions, we
see that $H(x)$ and its limit $h(x)$ have the same
functional form. To obtain the recurrence relations for
the series coefficients, first we divide the coefficients
(\ref{coef}) by $a$
and then take the limits of $\alpha_{n}^{(1)}/a$,
$\beta_{n}^{(1)}/a$ and $\gamma_{n}^{(1)}/a$
when $a\to \infty$, by using the relations (\ref{confluence}).
To permit the Leaver limit, the second step is given by the
substitutions
\begin{eqnarray}\label{thanks}
x=\frac{z_{0}-z}{z_{0}}, \ \ h(x)=e^{\rho z/(2z_{0})}U(z)
\end{eqnarray}
which convert Eq. (\ref{arscott}) into
\begin{eqnarray*}
z(z-z_{0})\frac{d^{2}U}{dz^{2}}+[-\delta z_{0}+(\gamma+\delta)z]
\frac{dU}{dz}+
\left[\frac{\gamma\rho}{2}-\sigma-
\left(\alpha-\frac{\gamma+\delta}{2} \right) \frac{\rho(z-z_{0})}{z_{0}}
-\frac{\rho^2z(z-z_{0})}{4z_{0}^2}\right]U=0.
\end{eqnarray*}
Then, comparing this with the GSWE (\ref{gswe}), we find that
\begin{eqnarray}\label{thanks2}
\alpha=i\eta+\frac{B_{2}}{2}, \delta=-\frac{B_{1}}{z_{0}},
\ \ \gamma=B_{2}+\frac{B_{1}}{z_{0}}, \ \ \rho=-2i\omega z_{0},
\ \ \sigma=-i\omega z_{0}\left(B_{2}+\frac{B_{1}}{z_{0}} \right)-B_{3}.
\end{eqnarray}
Therefore, thanks to Eq. (\ref{thanks}),
the solutions $U(z)$ for the GSWE (\ref{gswe}) are
obtained by writing
\begin{eqnarray}
U(B_{1},B_{2},B_{3};z_{0}, \omega,\eta;z)= e^{i\omega z}h\left( \alpha,\gamma,\delta, \rho,\sigma;
x=\frac{z_{0}-z}{z}\right)
\end{eqnarray}
and by inserting the parameters (\ref{thanks2})
into the right-hand side and also into the recurrence
relations obtained in the first step.

By this procedure, the solutions
$H_{1\nu}^{\{0,1\}}(x)$ and $H_{1\nu}^{\{a,\infty\}}(x)$
for the Heun equation give, respectively, the solutions $U_{1\nu}^{0}(z)$ and $\hat{U}_{1\nu}^{\infty}(z)$
for the GSWE, written in  Sec. VI .
To obtain solutions which lead to
the expansions $U_{1\nu}^{\infty}(z)$
and $\widetilde{U}_{1\nu}^{\infty}(z)$ for the GSWE,
we achieve an appropriate substitution
of the independent variable, followed by a transformation
of the dependent variable, such that the Heun equation
is transformed into another version of itself (with
a different set of parameters).
More precisely, from the transformation theory for the
Heun equation \cite{maier}, if
$H(a,q;\alpha, \beta,\gamma,\delta;x)$
is a solution of the Heun equation, then another
solution is obtained by means of (transformation
number 105 of Maier's table \cite{maier})
\begin{eqnarray}\label{maier105}
&&H(a,q;\alpha, \beta,\gamma,\delta;x)\to\nonumber\\
&&\left(x-1\right)^{-\alpha} H\left( a,q+\alpha(\alpha-\gamma-\delta+1)a;
\alpha, \alpha-\delta+1,\epsilon,\alpha-\beta+1;\frac{x-a}{x-1}\right).
\end{eqnarray}
Then, by identifying $H(a,q;\alpha, \beta,\gamma,\delta;x)$
with $H_{1}^{\{0,1\}}(x)$
and $\hat{H}_{1}^{\{a,\infty\}}(x)$
we find that
\begin{eqnarray}\begin{array}{l}
H_{1\nu}^{\{0,1\}}(x)\to\left( x-1\right)^{-\alpha}\displaystyle\sum_{n=-\infty}^{\infty}
b_{n}^{(1)}F\left( -n-\nu+\alpha-\frac{\gamma+\delta}{2},n+\nu+1+\alpha-
\frac{\gamma+\delta}{2};
\epsilon;\frac{x-a}{x-1}\right) ,
\vspace{3mm}\\
\hat{H}_{1\nu}^{\{a,\infty\}}(x)\to\left( x-1\right)^{-\alpha}
\displaystyle\sum_{n=-\infty}^{\infty}
c_{n}^{(1)}
\left[\frac{1-\frac{x}{a}}{x-1} \right] ^{-n-\nu-\alpha-1+\frac{\gamma+\delta}{2}}\times
\vspace{3mm}\\
\hspace{4cm}\widetilde{F}\left(n+\nu
+1+\alpha-\frac{\delta+\gamma}{2},n+\nu+1-\beta-\frac{\gamma+\delta}{2};
2n+2\nu+2;
\frac{x-1}{x-a}\right),
\end{array}
\end{eqnarray}
where
\begin{eqnarray}\label{dn}
c_{n}^{(1)}=(a)^{n}
\Gamma\left(n+\nu+1-\alpha-\frac{\gamma+\delta}{2} \right)
\Gamma\left(n+\nu+
1-\beta+\frac{\delta+\gamma}{2} \right)b_{n}^{(1)}.
\end{eqnarray}
Note that $\nu$ has not been transformed and that
the coefficients $b_{n}^{(1)}$ are the same which
appear in the original solutions.

Following the procedure outline above for $a\to0$ and using
the limits (\ref{limites}) for the hypergeometric functions
$F(a,b;c;y)$, we can show that
these give the solutions $U_{1\nu}^{\infty}(z)$
and $\widetilde{U}_{1\nu}^{\infty}(z)$ in
series of confluent hypergeometric
functions for the GSWE. In the first solution, before taking
the limit of the hypergeometric functions we make
the approximations
\[c=\epsilon=-\rho a, \ \ \frac{x-a}{x-1}=1+\frac{1-a}{x-1}\sim 1-\frac{c}{\rho(1-x)}.\]
In the second solution, first we must take
\[b\sim -\beta=\rho a, \ \ \frac{x-1}{x-a}\sim\frac{\rho(1-x)}{b}\]
and, in addition, we must find the recurrence relations
for the coefficients $c_{n}^{(1)}$ defined by Eq. (\ref{dn}).
However, to establish precisely this first set of solutions,
it is necessary to study the convergence of the two
last solutions and, then, apply  the so-called elementary
power transformations
of the dependent variable \cite{ronveaux} in order to
generate new sets of solutions. We have also to find
solutions valid for $a\in(0,1)$.

Finally we mention that the following series solution
for the Heun equation \cite{ronveaux}
\begin{eqnarray*}
H^{\{0,1\}}(x)=\displaystyle \sum_{n=0}^{\infty}d_{n}
x^{n},\ \ |a|>1,
\end{eqnarray*}
where the coefficients satisfy the relations
\begin{eqnarray*}
&&a(n+1)(n+\gamma)d_{n+1}-
\left[ an(n+\gamma+\delta-1)+
n(n+\alpha+\beta-\delta)+q\right]d_{n}\nonumber\\
\nonumber\\
&&+(n+\alpha-1)(n+\beta-1)d_{n-1}=0,
\end{eqnarray*}
converges for $|x|<|a|$ and
gives, by the above procedure, the solution
$U_{1}^{0}(z)$ of section II for the GSWE.
By analogy with the previous case, we may expect that the generalization of the solutions $U_{1}^{\infty}(z)$ and
$U_{1}(z)$ (in series of confluent hypergeometric functions)
are given by series of hypergeometric functions .

%
%
%


\begin{thebibliography}{99}
%
\bibitem{leaver1}E. W. Leaver,
J. Math. Phys. \textbf{27}, 1238 (1986).
%
\bibitem{humbert} P. Humbert, \textit{Fonctions de Lam\'e et Fonctions de
Mathieu}, M\'emorial des Sciences Math\'ematiques, X (Gauthier-Villards,
Paris, 1926).
%
\bibitem{ince}E. L. Ince, Proc. Lond. Math. Soc. {\bf 23}, 56
(1923).
%
\bibitem{eu}B. D. B. Figueiredo, J. Math. Phys.
{\bf 46}, 113503 (2005).
%
\bibitem{decarreau1}A. Decarreau, P. Maroni and
A. Robert,  Ann. Soc. Sci. Bruxelles, Ser. 1
{\bf T92}, 151 (1978) .
%
%
\bibitem{zaslavskii}O. B. Zaslavskii and V. V. Ulyanov,
{\it Sov. Phys. JETP} {\bf 60}, 991 (1984).
%
\bibitem{ulyanov}V. V. Ulyanov and O. B. Zaslavskii, {\it Phys. Rep.}
{\bf 261}, 179 (1992).
%
\bibitem{usheveridze1}A. G. Ushveridze, {\it Sov. J. Part. Nucl.}
{\bf 20}, 504 (1989).
%
\bibitem{usheveridze2}A. G. Ushveridze, {\it Quasi-Exactly Solvable
Models in Quantum Mechanics} (IOP, Bristol, 1994).
%
\bibitem{eu2}B. D. B. Figueiredo, J. Phys. A: Math. Gen. {\bf 35},
2877, (2002); J. Phys. A: Math. Gen. {\bf 35}, 4799
(2002) (corrigendum).
%
\bibitem{mano4}H. Suzuki. E. Takasugi and H. Hiroshi, {\it Prog. Theor. Phys.} {\bf 100}, 491 (1998).
%
\bibitem{mano5}H. Suzuki, E. Takasugi and H. Umetsu, {\it Prog. Theor. Phys.} {\bf 102}, 253 (1999).
%
\bibitem{olver}F. M. J. Olver, {\it Asymptotics
and Special Functions} (Academic Press, New York, 1974).
%
\bibitem{decarreau2}A. Decarreau, M. C. Dumont-Lepage,
P. Maroni, A. Robert and A. Ronveaux,
Ann. Soc. Sci. Bruxelles, Ser. 1 {\bf T92}, 53 (1978) .
%
\bibitem{ronveaux} {\it Heun's
Differential Equations}, edited by A. Ronveaux (Oxford University Press, 1995).
%
\bibitem{wilson}A. H. Wilson, {\it Proc. Roy. Soc. London}
{\bf A118}, 617 (1928).
%
\bibitem{schmidt}D. Schmidt and G. Wolf, {\it Double
Confluent Heun Equation}, Part C of \cite{ronveaux}.
%
\bibitem{mignemi}S. Mignemi, J. Math. Phys. {\bf 32},
3047 (1991).
%
\bibitem{malmendier}A. Malmendier,  J. Math. Phys. {\bf 44},
4308 (2003).
%
\bibitem{eguchi}T. Eguchi and A. J. Hanson, Phys. Lett. {\bf 74B},
249 (1978) .
%
\bibitem{arscott}F. M.  Arscott, Proc. Roy.
Soc. Edinburg {\bf A67}, 265 (1967) .
%
\bibitem{McLachlan}E. W. McLachlan,
{\it Theory and Application of Mathieu Functions}
(Dover, New York, 1964).
%
\bibitem{ince2}E. L. Ince, Proc. Lond. Math. Soc. {\bf 25}, 53 (1926) .
%
\bibitem{barber}W. G. Barber and H. R. Hass\'e,
Proc. Camb. Phil. Soc. {\bf 25}, 564 (1935 ).
%
\bibitem{otchik1}V. S. Otchik, in {\it Quantum Systems:
New Trends and Methods}, edited by A. O. Barut, I. D.
Feranchuk, Ya. M. Shnir and L. T. Tomil'chik
(World Scientific, 1995).
%
\bibitem{mano1}S. Mano, H. Suzuki and E. Takasugi, {\it Prog.
Theor. Phys.} {\bf 95}, 1079 (1996).
%
\bibitem{mano2}S. Mano, H. Suzuki and E. Takasugi, {\it Prog.
Theor. Phys.} {\bf 96}, 549 (1996).
%
\bibitem{mano3}S. Mano and E. Takasugi, {\it Prog. Theor. Phys.}
{\bf 97}, 213 (1997).
%
\bibitem{lindemann}F. Lindemann, Math. Ann.
{\bf 22}, 117 (1883).
%
\bibitem{watson}E. T. Whittaker and G. N. Watson,
{\it A Course of Modern Analysis}  (Cambridge
University Press, 1945).
%
\bibitem{fisher}E. Fisher,
Phil.Mag. {\bf 24}, 245 (1937).
%
\bibitem{bini}D. Bini, C. Cherubini, R. T. Jantzen,
B. Mashhoon,  Phys. Rev. {\bf D67},
084013 (2003).
%
\bibitem{gautschi}W. Gautschi, SIAM Rev.
{\bf 9}, 24  (1967).
%
\bibitem{arscott1}F. M. Arscott, {\it Periodic
Differential Equations} (Macmillan Company, New York, 1964).
%
\bibitem{erdelyi1}A. Erd\'elyi, W. Magnus, F. Oberhettinger and F. G. Tricomi (Bateman Manuscript Project)
{\it Higher Transcendental Functions},
vol. 1 (McGraw-Hill, New York, 1953).
%
\bibitem{abramowitz}{\it Handbook of Mathematical Functions},
edited by M. Abramowitz and I. A. Stegun (Dover, New York, 1965 ).
%
\bibitem{gradshteyn}I. S. Gradshteyn and I. M. Ryzhik,
{\it Table of Integrals, Series and Products}
(Academic Press, New York, 1994).
%
\bibitem{lemieux}A. Lemieux and A. K. Bose,
{\it Ann. Inst. Henri Poincar\'e} {\bf 10}, 259 (1969) .
%
\bibitem{luke}Y. L. Luke, {\it Integrals of Bessel functions}
(McGraw-Hill, New York, 1962).
%
\bibitem{scho1}E. Schr\"{o}dinger,
Commentationes Pontificiae Academiae Scientiarum,
{\bf 2}, 321 (1938).
%
\bibitem{scho2}E. Schr\"{o}dinger,  Proceedings of the
Royal Irish Academy {\bf A46}, 25 (1940).
%
 \bibitem{birrel} N. D. Birrell and P. C. W. Davies,
{\it Quantum Fields
in Curved Space} (Cambridge University Press, 1982).
 %
\bibitem{kleinman}C. J. Kleinman, Y. Hahn and
L. Spruch, Phys. Rev. {\bf 165}, 53 (1968).
%
\bibitem{buhring2}W. B\"{u}hring,  J. Math. Phys. {\bf 15}, 1451
(1974).
%
\bibitem{buhring1}W. B\"{u}hring, in
{\it Centennial Workshop on Heun's Equation}, edited by A. Seeger
and  W. Lay (Max-Plank-Institut
f\"ur Metallforchung, Institut f\"ur Physik, Stuttgart, 1990).
%
%
\bibitem{fackerell}E. D. Fackerell and R. G. Crossman,
{\it J. Math. Phys.} {\bf 18}, 1849 (1977).
%
\bibitem{erdelyi2}A. Erd\'elyi, {\it Q. J. Math.} {\bf 15}, 62 (1944).
%
\bibitem{heun}K. Heun,
{\it Math. Ann.} {\bf 33}, 161 (1899).
%
\bibitem{maier}R. S. Maier, e-print math.CA/0408317 v2 (2006).
%
\end{thebibliography}
\end{document}